\input epsf
\newif\iflanl
\openin 1 lanlmac
\ifeof 1 \lanlfalse \else \lanltrue \fi
\closein 1
\iflanl
    \input lanlmac
\else
    \message{[lanlmac not found - use harvmac instead}
    \input harvmac
    \fi
\noblackbox
\newif\ifhypertex
\ifx\hyperdef\UnDeFiNeD
    \hypertexfalse
    \message{[HYPERTEX MODE OFF}
    \def\hyperref#1#2#3#4{#4}
    \def\hyperdef#1#2#3#4{#4}
    \def\hypernoname{}
    \def\e@tf@ur#1{}
    \def\eprt#1{{#1}}
    \def\CERN{\centerline{CERN Theory Division}}
    \def\wl{W.\ Lerche}
\else
    \hypertextrue
    \message{[HYPERTEX MODE ON}
\def\eprt#1{{
#1}}
\def\CERN{\centerline{

CERN Theory Division}}
\def\wl{
 W.\ Lerche}
\fi
\def\weqn#1{\xdef #1{(\noexpand\hyperref{}%
{equation}{\secsym\the\meqno}%
{\secsym\the\meqno})}\eqno(\hyperdef\hypernoname{equation}%
{\secsym\the\meqno}{\secsym\the\meqno})\eqlabeL#1%
\writedef{#1\leftbracket#1}\global\advance\meqno by1}
\def\weqnalign#1{\xdef #1{\noexpand\hyperref{}{equation}%
{\secsym\the\meqno}{(\secsym\the\meqno)}}%
\writedef{#1\leftbracket#1}%
\hyperdef\hypernoname{equation}%
{\secsym\the\meqno}{\e@tf@ur#1}\eqlabeL{#1}%
\global\advance\meqno by1}
\def\abstract#1{
\vskip .5in\vfil\centerline
{\bf Abstract}\penalty1000
{{\smallskip\ifx\answ\bigans\leftskip 1pc \rightskip 1pc
\else\leftskip 1pc \rightskip 1pc\fi
\noindent \abstractfont  \baselineskip=12pt
{#1} \smallskip}}
\penalty-1000}
\def\cD{{\cal D}}

\def\ni{\noindent}
\def\tilde{\widetilde}
\def\bar{\overline}
\def\us#1{\underline{#1}}

\def\Coeff#1#2{{#1\over #2}}
\def\Coe#1.#2.{{#1\over #2}}
\def\coeff#1#2{\relax{\textstyle {#1 \over #2}}\displaystyle}
\def\coefff#1#2{\relax{\textstyle {#1 \over #2}}\displaystyle}
\def\coe#1.#2.{\relax{\textstyle {#1 \over #2}}\displaystyle}

\def\to{\rightarrow}
\def\notin{\hbox{{$\in$}\kern-.51em\hbox{/}}}

\def\del{\partial}

\def\a{a}
\def\b{b}
\def\lia#1{l^{(#1)}}

\def\xa(#1){n_{#1}}
\def\ss{\scriptstyle}

\def\aa(#1){\hfil\!\!{\ss #1}\!\!}
\def\aa(#1){\hfil{\ss #1}}

\def\xt(#1){\theta_{#1}}
\def\xz(#1){z_{#1}}
\def\someline{\ \ \vrule height 
3.14159265358979323846264338327950288419716939937510582097494459230781pt
width 
72.9243192144158235419816520453842473327310168044576374774186718458681pt
depth -1.5pt\ \ }

\def\frac#1#2{{#1\over#2}}

\def\frc#1#2{{\coeff1{#2}#1}}
\def\frcb#1#2{{\coeff1{#2}(#1)}}
\def\Frc#1#2{{\coefff1{#2}#1}}
\def\Frcb#1#2{{\coefff1{#2}(#1)}}
\def\doubref#1#2{\refs{{#1},{#2}}}
\def\hth/#1#2#3#4#5#6#7{{\tt hep-th/#1#2#3#4#5#6#7}}
\def\nup#1({Nucl.\ Phys.\ $\us {B#1}$\ (}
\def\plt#1({Phys.\ Lett.\ $\us  {B#1}$\ (}
\def\cmp#1({Comm.\ Math.\ Phys.\ $\us  {#1}$\ (}
\def\prp#1({Phys.\ Rep.\ $\us  {#1}$\ (}
\def\prl#1({Phys.\ Rev.\ Lett.\ $\us  {#1}$\ (}
\def\prv#1({Phys.\ Rev.\ $\us  {#1}$\ (}
\def\mpl#1({Mod.\ Phys.\ Let.\ $\us  {A#1}$\ (}
\def\atmp#1({Adv.\ Theor.\ Math.\ Phys.\ $\us  {#1}$\ (}
\def\ijmp#1({Int.\ J.\ Mod.\ Phys.\ $\us{A#1}$\ (}
\def\jhep#1({JHEP\ $\us {#1}$\ (}

\def\bb#1{{\bar{#1}}}
\def\bx#1{{\bf #1}}
\def\cx#1{{\cal #1}}
\def\tx#1{{\tilde{#1}}}

\def\rmx#1{{\rm #1}}
\def\us#1{\underline{#1}}
\def\fc#1#2{{#1\over #2}}
\def\frac#1#2{{#1\over #2}}

\def\br{\hfill\break}
\def\ni{\noindent}

\def\al{\alpha}\def\be{\beta}\def\ga{\gamma}
\def\om{\omega}
\def\p{\partial}
\def\Delop{\Delta_{bound}}
\def\lra#1{\matrix{#1\cr\longrightarrow\cr\phantom{1}}}

\def\CY{Calabi--Yau\ }

\def\IP{\bx P}\def\CC{\bx C}\def\ZZ{\bx Z}
\def\lm#1{l^{(#1)}}\def\tlm#1{\tilde{l}^{(#1)}}
\def\ss{\scriptstyle}

\def\Wtd{\cx W_{D=2}}
\def\Om{\Omega}
\def\zo{z_0}\def\ty{\tilde{y}}

\def\nihil#1{``#1"}
\lref\mlsm{P.~S.~Aspinwall, B.~R.~Greene and D.~R.~Morrison,
``Calabi-Yau moduli space, mirror manifolds and spacetime topology  change in string theory,''
Nucl.\ Phys.\ B {\bf 416}, 414 (1994), hep-th/9309097;\br
D.~R.~Morrison and M.~Ronen Plesser,
``Summing the instantons: Quantum cohomology and 
mirror symmetry in toric varieties,''
Nucl.\ Phys.\ B {\bf 440}, 279 (1995),
hep-th/9412236.}
\lref\wittop{
E.~Witten,
``On The Structure Of The Topological Phase Of Two-Dimensional Gravity,''
Nucl.\ Phys.\ B {\bf 340}, 281 (1990);\br
E.~Witten,
``Mirror manifolds and topological field theory,''
hep-th/9112056.}
\lref\modfcts{A.~Klemm, W.~Lerche and P.~Mayr,
``K3 Fibrations and heterotic type II string duality,''
Phys.\ Lett.\ B {\bf 357}, 313 (1995), hep-th/9506112;\br
V.~Kaplunovsky, J.~Louis and S.~Theisen,
``Aspects of duality in N=2 string vacua,''
Phys.\ Lett.\ B {\bf 357}, 71 (1995), hep-th/9506110;\br
B.~H.~Lian and S.~T.~Yau,
``Mirror maps, modular relations and hypergeometric series. II,''
Nucl.\ Phys.\ Proc.\ Suppl.\  {\bf 46}, 248 (1996), hep-th/9507153;\br
M.~Henningson and G.~W.~Moore,
``Counting Curves with Modular Forms,''
Nucl.\ Phys.\ B {\bf 472}, 518 (1996), hep-th/9602154.}
\lref\ft{S.~Ferrara, D.~L\"ust, A.~D.~Shapere and S.~Theisen,
``Modular Invariance In Supersymmetric Field Theories,''
Phys.\ Lett.\ B {\bf 225}, 363 (1989).}
\lref\tts{S.~Cecotti and C.~Vafa,
``Topological antitopological fusion,'' Nucl.\ Phys.\ B {\bf 367}, 359 (1991).}
\lref\cf{P.~Candelas and A.~Font,
``Duality between the webs of heterotic and type II vacua,''
Nucl.\ Phys.\ B {\bf 511}, 295 (1998), hep-th/9603170.}
\lref\oo{H.~Ooguri, Y.~Oz and Z.~Yin,
``D-branes on Calabi-Yau spaces and their mirrors,''
Nucl.\ Phys.\ B {\bf 477}, 407 (1996), hep-th/9606112.}
\lref\bkmv{
S.~Katz, P.~Mayr and C.~Vafa,
``Mirror symmetry and exact solution of 4D N = 2 gauge theories. I,''
Adv.\ Theor.\ Math.\ Phys.\  {\bf 1}, 53 (1998), hep-th/9706110;\br
P.~Berglund and P.~Mayr,
``Heterotic string/F-theory duality from mirror symmetry,''
Adv.\ Theor.\ Math.\ Phys.\  {\bf 2}, 1307 (1999), hep-th/9811217.}
\lref\Bat{V.V. Batyrev, {``Variations of the Mixed Hodge Structure
of Affine Hypersurfaces in Algebraic Tori"},
Duke Math. J. {\bf 69} (1993) 349.}
\lref\mirrorbook{
\nihil{Essays on mirror manifolds}, (S.\ Yau, ed.),
International Press 1992;\br
\nihil{Mirror symmetry II}, (B.\ Greene et al, eds.),
International Press 1997; plus references therein.}
\lref\govnew{S.~Govindarajan, T.~Jayaraman and T.~Sarkar,
``Disc instantons in linear sigma models,'' hep-th/0108234.}
\lref\asp{P.~Mayr,
``N = 1 mirror symmetry and open/closed string duality'', hep-th/0108229.}
\lref\agm{P.~S.~Aspinwall, B.~R.~Greene and D.~R.~Morrison,
``Measuring small distances in N=2 sigma models,''
Nucl.\ Phys.\ B {\bf 420}, 184 (1994), hep-th/9311042.}
\lref\wlsm{E.~Witten,
``Phases of N = 2 theories in two dimensions,''
Nucl.\ Phys.\ B {\bf 403}, 159 (1993), hep-th/9301042.}
\lref\bcov{M.~Bershadsky, S.~Cecotti, H.~Ooguri and C.~Vafa,
``Kodaira-Spencer theory of gravity and exact results for
quantum string amplitudes,''
Commun.\ Math.\ Phys.\  {\bf 165}, 311 (1994), hep-th/9309140;\br
I.~Antoniadis, E.~Gava, K.~S.~Narain and T.~R.~Taylor,
``Topological amplitudes in string theory,''
Nucl.\ Phys.\ B {\bf 413}, 162 (1994), hep-th/9307158.}
\lref\wittopop{E.~Witten,
``Chern-Simons gauge theory as a string theory,'' hep-th/9207094.}
\lref\ov{H.~Ooguri and C.~Vafa,
``Knot invariants and topological strings,''
Nucl.\ Phys.\ B {\bf 577}, 419 (2000), hep-th/9912123.}
\lref\vafaln{C.~Vafa,
``Superstrings and topological strings at large N,''
hep-th/0008142.}
\lref\hori{K.~Hori,
``Linear models of supersymmetric D-branes,'', hep-th/0012179.
}
\lref\vafaext{C.~Vafa,
``Extending mirror conjecture to Calabi-Yau with bundles,''
hep-th/9804131.}
\lref\ffsp{
P.~Mayr,
``Mirror symmetry, N = 1 superpotentials and tensionless strings on
Calabi-Yau four-folds,''
Nucl.\ Phys.\ B {\bf 494}, 489 (1997), hep-th/9610162.
}
\lref\HV{K.~Hori and C.~Vafa,``Mirror symmetry,'' hep-th/0002222.}
\lref\HIV{K.~Hori, A.~Iqbal and C.~Vafa,``D-branes and mirror
symmetry,''
hep-th/0005247.}
\lref\AViii{M.~Aganagic and C.~Vafa,
``Mirror symmetry and a G(2) flop,''
hep-th/0105225.}
\lref\AVii{M.~Aganagic, A.~Klemm and C.~Vafa,
``Disk instantons, mirror symmetry and the duality web,''
hep-th/0105045.}
\lref\AVi{M.~Aganagic and C.~Vafa,
``Mirror symmetry, D-branes and counting holomorphic discs,''
hep-th/0012041.}
\lref\AViv{M.~Aganagic and C.~Vafa,
``G(2) manifolds, mirror symmetry and geometric engineering,'' hep-th/0110171.}
\lref\Witbranes{
E.~Witten,
``Branes and the dynamics of {QCD},''
Nucl.\ Phys.\ B {\bf 507}, 658 (1997),
hep-th/9706109.}
\lref\vm{
M.~Marino and C.~Vafa,
``Framed knots at large N,''
hep-th/0108064.}
\lref\sheld{
S.~Kachru, S.~Katz, A.~E.~Lawrence and J.~McGreevy,
``Open string instantons and superpotentials,''
Phys.\ Rev.\ D {\bf 62}, 026001 (2000)
hep-th/9912151;\br
S.~Kachru, S.~Katz, A.~E.~Lawrence and J.~McGreevy,
``Mirror symmetry for open strings,''
Phys.\ Rev.\ D {\bf 62}, 126005 (2000)
hep-th/0006047;\br
S.~Katz and C.~M.~Liu,
``Enumerative Geometry of Stable Maps with Lagrangian Boundary
Conditions and Multiple Covers of the Disc,''
math.ag/0103074.}
\lref\quintic{
I.~Brunner, M.~R.~Douglas, A.~E.~Lawrence and C.~R\"omelsberger,
``D-branes on the quintic,''
JHEP {\bf 0008}, 015 (2000)
hep-th/9906200.}
\lref\zas{T.\ Graber and E.\ Zaslow,
\nihil{Open string Gomov-Witten invariants:
Calculations and a mirror `theorem',}
\eprt{hep-th/0109075}.
}
\lref\IK{A.\ Iqbal and A.\ K.\ Kashani-Poor,
\nihil{Discrete symmetries of the superpotential
and calculation of disk invariants,}
\eprt{hep-th/0109214}.
}
\lref\HKTY{S.\ Hosono, A.\ Klemm, S.\ Theisen and S.\ T.\ Yau,
\nihil{Mirror symmetry, mirror map and applications
to complete intersection Calabi-Yau spaces,}
\eprt{hep-th/9406055}.}
\lref\hktyI{
S.~Hosono, A.~Klemm, S.~Theisen and S.~T.~Yau,
``Mirror symmetry, mirror map and applications to Calabi-Yau hypersurfaces,''
Commun.\ Math.\ Phys.\  {\bf 167}, 301 (1995), hep-th/9308122.
}
\lref\morr{D.R. Morrison,``Compactifications of moduli spaces 
inspired by mirror symmetry'', alg-geom/9304007.}

\vskip-2cm
\Title{\vbox{
\rightline{\vbox{\baselineskip12pt\hbox{CERN-TH/2001-301}
                            \hbox{hep-th/0111113}}}\vskip.5cm}}
{On $\cx N=1$ Mirror Symmetry for Open Type II Strings}

\abstractfont
\vskip 0.8cm
\centerline{\wl\ and P. Mayr}
\vskip 0.8cm
\centerline{\CERN}
\centerline{CH-1211 Geneva 23}
\centerline{Switzerland}
\vskip 0.3cm
\abstract{%
We study the open string extension of the mirror map for
$\cx N=1$ supersymmetric type II vacua with D-branes
on non-compact Calabi-Yau manifolds. Its definition is given 
in terms of a system of differential equations that
annihilate certain period and chain integrals.
The solutions describe the flat coordinates on
the $\cx N=1$ parameter space, and the exact, disc 
instanton corrected superpotential on the
D-brane world-volume. A gauged linear sigma model
for the combined open-closed string system is also given.
It allows to use methods of toric geometry to describe
D-brane phase transitions and the $\cx N=1$ K\"ahler cone. 
Applications to a variety of D-brane geometries are described 
in some detail.
}
\vskip1cm
\Date{\vbox{\hbox{ {November 2001}}
}}
\goodbreak

\parskip=4pt plus 15pt minus 1pt
\baselineskip=14pt
\leftskip=8pt \rightskip=10pt
%
\newsec{Introduction}
Closed string mirror symmetry \mirrorbook\ has been very effective in
exactly computing instanton effects in $\cx N=2$ supersymmetric
type II strings. An important aspect is that certain holomorphic
quantities in the physical type II string theory are
computed by the topologically twisted theory \wittop, 
which provides a simplified structure essential for the
application of mirror symmetry. 

There is also an open string version of the topologically
twisted string \wittopop, which computes amplitudes of physical
type II open-closed string systems with $\cx N=1$
supersymmetry in four dimensions
\bcov-$\phantom{\ov}$\hskip-10pt\vafaln.
Specifically, one version of it, the A-model, 
is related to the type IIA string compactified on a 
\CY (CY) 3-fold $Y^*$, with D6-branes partially wrapping 3-cycles 
in $Y^*$ and filling space-time. Mirror symmetry relates
the A-model to the B-model, which computes amplitudes of 
the type IIB string with odd-dimensional branes wrapping
holomorphic sub-manifolds in the mirror manifold $Y$ of $Y^*$.
An extension of mirror symmetry that relates the topological
amplitudes of a mirror pair of D-brane geometries 
has been pioneered \refs{\vafaext,\AVi} and leads
to an expression for the exact instanton corrected superpotential
for a class of non-compact D6-branes \AVi. See also
{\quintic$\phantom{\AVii\sheld\AViii\asp\govnew\zas\IK}$\hskip-110pt-\AViv}\
for other papers on this subject.

Based on the developments of \AVi, a definition of
$\cx N=1$ open string mirror symmetry very similar to that for
closed strings has been given in \asp. 
A system of differential equations has been derived, whose
solutions around a limit point of maximal unipotent
monodromy describe the mirror map for the flat
coordinates of the combined open-closed string
moduli space\foot{We will loosely refer to the complex
scalar manifold of vev's of the massless chiral
$\cx N=1$ multiplets as the moduli space, although
there is a superpotential for them which
will fix some of these vev's. The notation ``moduli''
is partially justified by the fact that in the
appropriate regime the perturbative
superpotential is zero.}. The remaining solutions
represent the exact instanton sum for the
non-perturbatively generated $\cx N=1$ superpotential.
It was also argued that this differential structure describes
the restricted geometry of the holomorphic F-terms,
referred to as $\cx N=1$ special geometry in the
following. The wording reflects the fact that
this geometry is a close relative of 
$\cx N=2$ special geometry, however defined
by a set of {\it several} independent holomorphic functions,
as opposed to the single holomorphic prepotential for $\cx N=2$.

In this note we develop these ideas further and
generalize them in various ways. In particular
the arguments in \asp\ have been based on a new duality
between the open-closed type II strings in four dimensions
and a closed string background without D-branes in two dimensions.
Instead we derive in section 2 the differential
system for the holomorphic $\cx N=1$ amplitudes 
directly in the B-model for the four-dimensional open-closed string
theory. Although the open/closed string duality is an
interesting subject in itself and seems to be quite
a general phenomenon, our derivation here provides
the alluded to differential system for $\cx N=1$ mirror symmetry
independently of the existence of a closed string dual.
We proceed with a study of the general properties of the 
solutions. 
In section~3 we describe the construction of
a gauged linear sigma model (LSM) for the A-model that represents the
mirror D-brane geometry. We use the techniques of
toric geometry to define the K\"ahler cones of the
$\cx N=1$ moduli space and study phase transitions between
different classical vacua.
In section~4 we discuss some aspects of $\cx N=1$ special geometry
and the so-called framing ambiguity.
In the Appendix we included detailed computations for 
D-branes on a collection of non-compact CY 3-folds.

\newsec{B-model: Picard-Fuchs equations for open strings}
Our first aim will be to study the $\cx N=1$ moduli space of the 
B-model with D-branes and to derive a system of differential 
equations for it. The solutions of this generalized 
Picard-Fuchs (PF) system 
describes the flat coordinates on this space, as well as
the holomorphic topological correlation functions on it.
The differential system will agree with that derived in \asp\ for the
cases that have closed string duals. A 
differential constraint on the superpotential, which is however not
equivalent to the PF system described below, has recently been 
proposed in \govnew.

\subsec{Generalized Picard-Fuchs equations for open strings}
The ordinary Picard-Fuchs equation is a differential equation that
expresses the linear dependence of $k+1$ $p$-forms in the cohomology
group $H^p(Y,\CC)$ of dimension $h^p(Y)=k$. E.g. for
a CY 3-fold $Y$ with\foot{$h^{p,q}(Y)=\rmx{dim}H^{p,q}(Y)$.} 
$h^{1,2}(Y)=1$, there is a differential equation for the holomorphic
3-form $\Om(z)$ 
$$
\cx D\, \Om(z) = \sum_{i=0}^4 f_i(z) \fc{d^i}{dz^i}\, \Om(z) = d\eta(z).
$$
Here $z$ denotes the single complex structure modulus. The above
equation expresses the fact that the sum of the
five 3-forms $\fc{d^i}{dz^i}\Om$ for $i=0,...,4$ must be 
linearly dependent in the cohomology group $H^3(Y,\CC)$ and thus is 
proportional to an exact form $d\eta$. 
It follows that the differential operator $\cx D$ annihilates 
the period integrals,  
$\cx D \int_{\ga}\Om=0$ where $\ga$ is
a topological ($z$-independent)  3-cycle in $H_3(Y,\bx Z)$.

The open-string sector of the B-model 
consists of D5-branes wrapped on a 2-cycle $C$ in $Y$.
The superpotential for the D5-brane on the curve $C$ is 
\doubref\AVi\Witbranes

\eqn\chainsp{
W(\zo)=\int_{_{\Gamma(\zo)}}\hskip-9pt \Om=\int_{_{C(\zo)-C_*}}\hskip-15pt\om
\quad ,
}
where $\Gamma$ is a 3-chain with boundary $C-C_*$ and $\Om$ may be
locally written as $\Om = d\omega$ on $\Gamma$. Moreover $z_0$ is
an open string modulus that moves the position of $C$ in $Y$ and 
$C_*$ is a fixed reference curve homologous to $C$. 
The Picard-Fuchs operator applied to the above chain integral
gives a non-zero result from the boundary 
\eqn\pfii{
\cx D \int_{\Gamma} \Om(z)=\int_{\Gamma}d\eta(z)=\int_{\p\Gamma}\eta(z).
} 
In fact the non-vanishing term on the r.h.s. may be interpreted
as a new 
observable in the open string sector for each boundary component
$c_i\subset \p \Gamma$. This is closely related to the fact that in 2d terms
the BRST variation of a closed string bulk observable 
is zero up to a boundary term 
proportional to an open string observable.

Note that, contrary to the closed string periods, the chain integral 
$\chainsp$ depends in particular on the choice of a representative for
the cohomology class $\Om \in H^3(X,\CC)$. In other words, {\it the
exact piece of $\Om$ is observable in the open string sector}.
Specifically,
the exact piece of $\Om$ corresponds to a quantum number of the 
open-closed string vacuum which needs to be fixed to define
the system. As will be discussed somewhere else it is integrally
quantized and related to the framing ambiguity discussed in 
refs.\refs{\AVii,\vm}.

The equation \pfii\ suggests that an appropriate modification
of the Picard-Fuchs operator $\cx D$ may annihilate the 
non-vanishing boundary term. 
To describe the necessary generalization, we recall briefly the 
construction of the complex 
structure moduli space of $Y$.
Concretely we study the topologically twisted
B-model on a toric CY manifold~$Y$ defined as 
the vacuum geometry of a 2d linear sigma model \wlsm\
with superpotential
\eqn\mirrorsp{
\Wtd=\sum_{i=1}^{N+3} a_i \ty_i,
\qquad \prod_i \ty_i^{\lm a_i}=1,\qquad a=1,...,N.}
Here the $N+3$ variables $\ty_i$ take 
value in $\CC^*$ and the $a_i$ are
constants that parametrize the complex structure. A concrete
CY is specified by the integral charge vectors $\lm a$
that define the set of $N=h^{1,2}(Y)$ 
relations on the r.h.s of \mirrorsp.
After solving the relations for a choice of three coordinates
$\ty_{i_1},\ty_{i_2},\ty_{i_3}$, one obtains the superpotential 
$\Wtd(\ty_{i_1},\ty_{i_2},\ty_{i_3};a_i)$.
An equivalent definition \HIV\ is in terms of the hypersurface
\eqn\tdspii{P=F(y_{i_1},y_{i_2};a_i)+xz=0,\qquad y_{i_k}=
\fc{\ty_{i_k}}{\ty_{i_3}},}
where $F=\cx W/\ty_{i_3}$. 

In the following we will often set the indices $\{i_n\}$ 
to the specific values $(1,2,3)$ to simplify notation. 
This choice of coordinates is appropriate to describe
D-branes classically located in a patch where $\ty_3$ is ``large''. D-branes
in other patches may be described similarly after an appropriate relabeling
of the coordinates. We will switch back to a global notation 
with general $i_n$ where it is useful.

The naive moduli space $\cx M_0$ is 
$$
\cx M_0= (\CC^*)^{M}/(\CC^*)^{m},
$$ 
where $(\CC^*)^M$ is parametrized by the $a_i$ 
and the quotient by $(\CC^*)^m$ is generated by the reparametrizations
\eqn\scalings{
\ty_{i} \to \lambda_{i} \ty_{i},\qquad \lambda_{i} \in \CC^*,}
of the independent coordinates $\ty_i$. In the present case, $M=N+3$ and $m=3$.
The true moduli space is obtained from $\cx M_0$ by an appropriate
compactification that adds some limit points and subsequently 
removing the discriminant locus
where the surface $Y$ is singular \agm. The first step is accomplished by
passing to the $(\CC^*)^m$ invariant coordinates 
\eqn\defz{z_a=\prod_i a^{\lm a_i}_i,\qquad a=1,...,N,}
which provide good local coordinates for the complex structure moduli
space $M$ of $Y$ in a neighborhood of $z_a=0$.

The complex structure of $Y$ is locally also 
parametrized by its periods $\int_{\ga_\al}\Om$, where $\ga_\al\in H_3(Y,\ZZ)$.
The periods satisfy a system of differential equations of generalized 
hyper-geometric, so-called GKZ type \Bat.
It is defined by two sets of differential operators. The first set is
of the form
\eqn\gkzi{
\tilde{\cx D}_j=\sum_i \nu_{i,j}\, a_i \fc{\p}{\p a_i} - \beta_j,}
where $\beta_j$ is a vector of exponents which is identically
zero in the non-compact case. Moreover the $\nu_i$ are $N+3$
vertices of the polyhedron $\Delta$ defining the toric variety $Y$; its
construction will be described in the next section.
For the moment it suffices to know that the differential operators
$\tx {\cal D}_j$  express the invariance of the period integrals
under the $\CC^*$ scalings \scalings. One may therefore solve the
equations \gkzi\ by using the $\CC^*$ scalings to write the
3-form $\Om(a_i)$ as a function $\Om(z_a)$ of the $z_a$, only.\foot{In
the compact case there is also a normalization factor corresponding
to non-zero exponents $\beta_j$.}

On the surface $P=0$, the 3-form $\Om$ is given by the residuum
formula 
\eqn\holtf{
\Om=\fc{dy_1}{y_1}\fc{dy_2}{y_2}\fc{dx\, dz}{P}
\ \lra{\scriptstyle res} \
\fc{dy_1}{y_1}\fc{dy_2}{y_2}\fc{dz}{z}.
}
{From} this explicit form it is easy to see that 
$\Om$ is annihilated by a second set of differential 
operators of the form
\eqn\defGKZ{
\cx D_{a}=\prod_{\lm a _i>0}\big(\fc{\p}{\p a_i}\big)^{\lm a _i}-
\prod_{\lm a _i<0}\big(\fc{\p}{\p a_i}\big)^{-\lm a _i}.
}
These operators 
reflect the relations between the monomials $\ty_i$ in \mirrorsp.
They also annihilate the period integrals of $\Om$ on 
a basis of topological cycles $\gamma_a$ spanning $H_3(Y,\ZZ)$.

To describe the open string modifications
to the above picture 
we consider from now on a specific class of D5-branes on 
a non-compact 2-cycle $C$ as in \AVi. These are mirror to 
D6-branes on the mirror manifold $Y^*$ with a non-zero,
however entirely non-perturbative superpotential $W$.\foot{%
Supersymmetric D-branes and their mirrors 
have been described in \refs{\oo,\HIV,\AVi,\hori}.}
Specifically, the curve $C$ is defined by the equations
\eqn\defc{
x=F=0,\qquad y_i=y_i(z),\qquad y_i(\infty)=y_i^*.
}
The reference curve $C_*$ may be chosen to be the holomorphic 
cycle $y_i(z)=y_i^*$. The rational equivalence class of $C$ is 
parametrized holomorphically by the quantity
\eqn\defbound{
y_1(0)=\zo,}
which measures the deformation from the configuration
$C_*$. Note that the value of $y_1$ fixes the value of $y_2$
by the constraint $F=0$. With an appropriate labeling of coordinates,
any family of 2-cycles $C$ may be written as \defbound; alternatively
one may write the general boundary condition as 
\eqn\defboundii{\ty_i(0)=z_0\, \ty_j(0),} which
will be also be needed below and reduces to \defbound\ for
the coordinates $(i,j)=(1,3)$ used in the definition \tdspii.

An important fact is that the
open string sector -- the location of the D-brane on the 
curve $C$ -- breaks the $\CC^*$ scaling symmetry \scalings\ of the 
closed string sector. Specifically,
the $\CC^*$ symmetry $y_1\to \lambda y_1$ does not leave invariant 
the boundary condition \defbound. Instead one may 
use this $\CC^*$ action to move the moduli dependence 
from the integration contour to the integrand
\eqn\cint{
\int_{\Gamma(\zo)} \Om(z_a)=\int_{\Gamma} \Om(z_a;\zo).
}
Here $\Gamma=\Gamma(1)$ is a {\it topological} 3-chain\foot{A
rigorous mathematical definition exists.}
independent of the open string modulus. Moreover $\Om(z_a,\zo)$ 
is obtained from $\Om(z_a)$ by rescaling of $y_1$, which 
replaces  $P(y_1,y_2)$ with $P(\zo \, y_1,y_2)$ in the 
definition \holtf. 
The 3-form $\Om(z_a;\zo)$ and its integrals over
topological 3-cycles {\it and} the 3-chain $\Gamma$ is annihilated by the GKZ 
operators $\cx D_a$, rewritten in terms of  $z_a$ and the new modulus $\zo$.

The same integrals are in addition annihilated by a further
``boundary differential operator'' $\cx D_0$, as we will show now.
The combined system of differential operators $\{D_0,D_a\}$ will
be complete in the sense that the integrals $\int_{\ga_a}\Om$ 
and $\int_\Gamma \Om$ provide a basis of solutions. 

To derive the additional differential operator satisfied
by the relevant integrals of $\Om$, consider a hyperplane $H\in Y$
that intersects $\p\Gamma$ in the point \defbound, or more generally
\defboundii
\eqn\newrel{ 
\vbox{\offinterlineskip\tabskip=0pt\halign{
\strut 
\hfil$#$~& $#$~\hfil& $#$~\hfil&$#$~\hfil&$#$~\hfil
\cr
H:\ \ z_0^{-1}\, \prod_{i} \ty_i^{\, \lm 0_i}=1,\qquad
\lm {0}&=(0,...,&,1,...&,-1,...&,0)\ .\cr
& &{\scriptstyle i-th}& {\scriptstyle j-th\  position}\cr
}}
}
We define a 2-form $\Omega_0$ by $\Omega=\Omega_0\fc{dy_1}{y_1}$.
The restriction of $\Omega_0$ to $H$ 
is annihilated by the differential operator
$$
\cx L\, \Om_0(z_a,\zo)|_H=0\, \qquad \cx L=\p_{a_i}-\p_{a_j}.
$$
Note that the origin of $\cx L$ is very similar to that of the 
operators $D_a$ in \defGKZ; in fact the differential 
operators $\{\cx D_a,\cx L\}$ correspond to the GKZ operators 
\defGKZ\ derived from the ``boundary superpotential'' 
$$
\cx W_{bound}\ =\ 
\sum_{i=1}^{N+3} a_i \ty_i,
\qquad \prod_i \ty_i^{\lm \al_i}=1,\qquad \al=0,...,N,
\weqn\boundpot
$$
which depends, as compared to the closed string ``bulk'' superpotential
\mirrorsp, also on the additional D-brane modulus 
$z_0=\prod a_i^{\lm 0_i}$\foot{A CFT interpretation of $\cx W_{bound}$
exists, as will be discussed elsewhere.}.

The operator $\cx L$ does neither annihilate the periods 
nor the chain integrals. To derive a differential operator $\cx D_0$ 
from it that does so, consider an infinitesimal 
variation of the chain integral 
under a shift $\zo \to \zo + \delta \zo$ 
$$\eqalign{
&\int_{_{\Gamma(\zo+\delta \zo)}}\hskip-20pt \Om(z_a)
-\int_{_{\Gamma(\zo)}}\hskip-10pt\Om(z_a)=
\int_{\zo}^{\zo +\delta \zo} \Big(\int \Om_0(z_a) \Big) \fc{dy_1}{y_1}=
\phantom{\matrix{1\cr1\cr1\cr1\cr}}\cr
&\int_{1}^{1 +\delta \zo/\zo} \Big(\int \Om_0(z_a;\zo) \Big) \fc{dy_1}{y_1}=
\big(\int \Om_0(z_a;\zo)\big)|_H\ \cdot\  \delta \zo/\zo.
}$$
Thus the searched for differential operator is 
$$
\cx D_0\, \int_{\Gamma(\zo)}\Om(z_a) =0,
\qquad \cx D_0=\cx L\, \zo\fc{d}{d\zo},
$$
which annihilates trivially also the period integrals 
$\int_{\ga_a}\Om$ since they do not depend on $z_0$. The
above derivation may also be generalized to yield
a complete system for several independent chain integrals 
parametrized by several open string  moduli $z_0^\mu,\,
\mu=1,...,l$.

It remains to rewrite the 
system of differential operators $\{\cx D_\al\}=\{\cx D_a,\cx D_0\}$ in 
terms of derivatives of the good local coordinates $(z_a;z_0)$.
The result can be given in a simple closed form as follows. 
Define an 
extended set of charge vectors
$\lm \al, \al=0,...,N= h^{1,2}$, with
\def\lmh#1{\tilde{l}^{(#1)}}
\vskip-18pt
\vbox{
\vskip 10pt
\eqn\defmoris{
\vbox{\offinterlineskip\tabskip=0pt\halign{
\strut
\hfil$#$~
&$#$~\hfil
&\hskip 5pt$#$\hskip 5pt
&\hskip5pt$#$\hskip10pt
&\hfil$#$~&\hfil$#$~&\hfil$#$~&\ \ \ $#$\cr
\lmh {a}&=(&&\lm a(Y)&;&0&0\, ),&\ \ a=1,...,N,\cr
\lmh {0}&=(0,&...,0,1,0,...&...,0,-1,0,...&...,0\, ;&1&-1\, ).&\cr
&&{\scriptstyle i-th\ position}&\ {\scriptstyle j-th\ position}\cr
}}}}
\vskip-15pt\ni
With these definitions, the extended system of differential 
operators can be written in the closed form 
\def\lmh#1{\tilde{l}^{(#1)}}
\eqn\defGKZii{
\cx D_\al=
\prod_{\lmh \al_i>0}\prod_{j=0}^{\lmh \al_i-1}
\big(\sum_\be \lmh \be_i\theta_\be-j \, \big)-
z_\al\, 
\prod_{\lmh \al_i<0}\prod_{j=0}^{-\lmh \al_i-1}
\big(\sum_\be \lmh \be_i\theta_\be-j\, \big).}
A system of differential equations of the type 
\defmoris, \defGKZii\ has been derived in \asp\
using a duality of the open-closed string background on $Y$ to a 
closed string background without branes on a CY 4-fold. Here
we have obtained the differential equations that govern the 
open-closed holomorphic data purely from open string arguments.
In particular the above line of arguments is applicable no matter
whether a closed string dual exists. Note that 
a given set of charge vectors $\lmh a$ is in general
not equivalent to specifying a toric manifold $X$. It would be 
interesting to know under which conditions the open-closed
string data define a toric manifold for a closed string background.

\subsec{Open string mirror map and $N=1$ superpotentials}
We proceed with a discussion of  
some general properties of the solutions to 
the open-closed string GKZ system \defGKZii.
As for notation, we drop the tildes on $\tlm \al$ and take 
the following convention in this section\foot{In general 
there will be several
patches in the moduli that include different large complex structure 
limit points. For the present section,
the notation is adapted to an ``outer'' phase as defined in the
next section.}:
\eqn\mori{\eqalign{
\lm a&=(\lm a(Y)\hskip20pt ;0,\ \, 0),\qquad a=1,...,N= h^{1,2},
\cr
\lm 0&=(1,-1,0,...0;1,-1).
\cr
}}
The first entry in each row corresponds to 
the non-compact direction and is labelled by $\lm\al_0$.

For the above choice of $\lm \al$,
the algebraic coordinates \defz\ will be good coordinates
near the point $z_\al=0$ of so-called 
maximally unipotent monodromy.
It is quite non-trivial that this concept generalizes to include 
the open string moduli and this is in fact 
a consequence of the underlying $\cx N=1$ special geometry\foot{See
\morr\hktyI\ for the selection of the limit point in the closed string case.}.
A concrete justification can be given by constructing the appropriate
gauged linear sigma model for the A-model on the mirror manifold, 
as will be done in the next section.

The behavior of the solutions $\om_i$ near the limit point $z_\al=0$ 
is of the form $\om_i\sim (1;\ln(z);\ln(z)^2,...)$, reflecting the 
distinct monodromy of the solution vector $\om_i$ on
a circle around $z_\al=0$. The solutions may be 
defined in terms of the power series $\om=\sum_{n_\al}c(n_\al;\rho_\al)\,
\prod_\al z_\al^{n_\al+\rho_\al}$, with 
$$
c(n_\al;\rho_\al)=
\fc{1}
{\prod_i\Gamma(1+\sum_\al \lm a_i(n_\al+\rho_\al)\, )}\ ,
$$
and derivatives of $\om$ with respect to the indices $\rho_a$.
The series solution $\om|_{\rho_\al=0}$ is in fact 
a constant as a consequence of the non-compactness of the 3-fold. 
The solutions $\fc{\p}{\p\rho_\al}\, \om|_{\rho_\al=0}$ with
single logarithmic behavior define the flat coordinates of 
the open-closed topological string theory \asp. 
They have the form 
\eqn\flatco{\eqalign{
t_\al(z_\be)\ &=\ \coeff1{2\pi i}\log(z_\al)+S_\al(z_\be)\ ,\cr
}}
where $t_\al,t_0$ are the closed and open string flat coordinates. 
A general property is that the series
\eqn\mirrorser{
S_\al(z_\b)\ =\ \sum_{n_\b\geq0} c_\al(n_\b) {z_1}^{n_1}\dots {z_{N}}^{n_{N}}
}
depend only on the bulk moduli and expressly 
not on the boundary modulus $\zo$ \refs{\AVii,\asp}. 
The coefficients can be written as 
\eqn\can{
c_\a(n_\b)\ =\ (-)^{m_k+1}\lia\a_k(m_k-1)!
\Big(\prod_{i\not=k}(\sum_\b\lia\b_in_\b)!\Big)^{-1}\ ,
}
where $m_k$ is such that $\sum_\b\lia\b_in_\b\equiv -m_k<0$.
It can be shown that $c_\a(n_\b)$ is non-zero only when there
exists a single such positive $m_k$.
Moreover, the series for the D-brane modulus mirror map is a simple linear
combination of the bulk moduli series \asp:
\eqn\linco{
S_0(z_\a)\ =\ \sum (A^{-1})_\b^\a A_0^\a S_\b(z_\a)\ ,
}
where $A$ is the integer coefficient matrix characterizing the linear
part of the PF system, ie., $\cD_\a=z_\a\sum A^\a_\b\theta_\b+...$ (when
$A$ is degenerate one needs to restrict to linearly independent
$S_\a(z_\b)$).

The superpotential corresponds to the $\zo$ dependent
double-logarithmic solution corresponding to a double
derivative of $\om$ w.r.t. the indices.
More precisely, it is obtained by dropping the
logarithmic pieces,  because the classical superpotential for 
the D-brane is zero by construction. We find:
\eqn\Wouter{
W(z_\a,z_0) = \!\!
\sum_{{n_\a\geq0,n_0>0\atop n_0-\sum\lia\a_1n_\a>0}}
\!\!
{(-)^{\sum\lia\a_1n_\a}(n_0-\sum\lia\a_1n_\a-1)!
\over 
n_0 (n_0+\sum\lia\a_0n_\a)!
\prod_{i=2}^{N+2}(\sum\lia\a_in_\a)!}{z_0}^{n_0} 
\prod_{\a}{z_\a}^{n_\a}
\ .
}
This result was obtained by applying the Frobenius method  (as
nicely explained in \HKTY) and noting that there can be a contribution if and
only if two $m_k$ are positive. Moreover it turns out that these 
must be $m_1$ and $m_{N+4}$ in order to give actual solutions of the Picard-Fuchs system \defGKZii.
Note that the purely $z_0$ dependent terms give the dilogarithm
function,
\eqn\dilog{
W(0,z_0)\ =\ \sum_{n_0>0}{1\over {n_0}^2}\,{z_0}^{n_0}\ \equiv\ 
{\rm Li}_2(z_0)\ ,
}
which is consistent in that the large radius limit reproduces the 
known result for $\CC^3$~\AVi.

Inserting the inverse mirror map $z_{\al}(t_\be)$  \flatco\ into
$W(z_\al)$ one obtains the expansion of the superpotential in terms of the
exponentiated flat coordinates $q_\al\equiv e^{2\pi it_\al}$. It 
is predicted to be of the form \ov:
\eqn\Wpot{
W(q_\a,q_0)\ =\ \sum_{n_\a,n_0} N_{n_\a,n_0}\,
{\rm Li}_2({q_0}^{n_0}\prod_\a {q_\a}^{n_\a})\ .
}
Here the {\it integral} coefficients $N_{n_\a,n_0}$ count the numbers%
\foot{The definition of the integers $N_{n_\a,n_0}$
is  motivated by physics and 
based on appropriately counting the numbers of D4-brane
domain walls wrapping the discs \AVi\ov.} of disc instantons 
of the corresponding degrees.

We can improve the formula \Wouter\ by noting that
it can be conveniently summed over $z_0$.
Defining $\xi(n)={\rm max}(\sum\lia\a_1n_\a,0)$, we can rewrite the
superpotential as
\eqn\Whyp{\eqalign{
W(z_\a,z_0) &= \!\! \sum_{n_\a\geq0}
{(-)^{\sum\lia\a_1n_\a} (\xi(n)-\sum\lia\a_1n_\a)!
\over
(1+\xi(n))
\Gamma(2+\xi(n)+\sum\lia\a_0n_\a)
\prod_{i=2}^{N+2}(\sum\lia\a_in_\a)!}\cr\cr
\times\ \
&{}_3F_2\left(
{1, \qquad 1+\xi(n),\qquad 1+\xi(n)-\sum\lia\a_1n_\a
\atop
2+\xi(n),\ 2+\xi(n)+\sum\lia\a_0n_\a}
;\, z_0\right){z_0}^{1+\xi(n)} 
\prod_{\a}{z_\a}^{n_\a}.}}
\ni
This form facilitates analytic continuation in the boundary modulus $z_0$,
and in particular 
allows to determine the $n_0$ dependence of the instanton 
coefficients in a closed form. This is because
the hyper-geometric function $_3F_2$ in \Whyp\ has 
degenerate arguments so that it takes the following
generic functional form: $c \log(1-z_0)+(1-z_0)^{-d}Q(z_0)$,
where $c,d$ are constants and $Q(z_0)$ is a {\it finite} polynomial.
That this polynomial is finite translates to the property
of the generating function:
\eqn\Nx{
N_{n_\a}(x)\ \equiv\ \sum_{n_0} N_{n_\a,n_0} x^{n_0}
}
to be a ratio of two finite polynomials in $x$; 
examples for such generating functions are presented in 
Appendix~A. Moreover note that upon taking a derivative,
\Whyp\ can be further condensed because
the ordinary hyper-geometric function on the r.h.s.\ of
$$
z_0\,\del_{z_0}
\left[{z_0}^{1+\xi(n)}{}_3F_2\left(\dots;\, z_0\right)\right]
=
{{z_0}^{1\!+\!\xi(n)}\over\xi(n)!}{}_2F_1\left(1,\,1\!+\!\xi(n)\!-\!
\sum\lia\a_1n_\a,\,2\!+\!\xi(n)\!+\!\sum\lia\a_0n_\a;\,z_0\right)
\weqn\hypid
$$
has a very simple structure. 
For example, for $\sum\lia\a_1n_\a=\xi>0$ and 
$\sum\lia\a_0n_\a+\sum\lia\a_1n_\a\equiv -k\leq0$, it reduces to
$$
{{z_0}\over \Gamma(2-k)}\,{}_2F_1(1,1,2-k;\,z_0)\ =\ 
\cases{
 -\log(1-z_0), & $k=0$ \cr
(k-1)!\big({z_0\over 1-z_0}\big)^k, & $k\geq1$.\cr
}
\weqn\ftwosimp
$$
These simple expressions lead to a 
firm control over the moduli sub-space 
spanned by $z_0$. 

\newsec{A-model: a gauged LSM for open strings and topology changing phase
transitions}
As promised in \asp\ and the previous section, we complete
in the following the toric description of the D-brane geometry
by constructing a gauged LSM for it. It describes the 
A-model for the mirror D6-branes wrapped on special Lagrangian (sL)
3-cycles on the mirror manifold $Y^*$. This will in particular be 
useful to define the phases of the $\cx N=1$ moduli space and
the K\"ahler cones for it. We will also consider D-brane 
phase transitions which change the topology of the gauged LSM.
The LSM can be interpreted as describing a CY 4-fold $X^*$
for a closed string compactification mirror to those considered in \asp.
Alternatively there could be an interpretation
in terms of a ``boundary linear sigma model''. This
point of view has been pursued in \govnew, although with 
results different from ours\foot{The LSM considered 
in \govnew\ does not obviously reproduce the relevant PF system \defGKZii.}.

The connection between this gauged LSM and the D-branes 
on $Y^*$ has been demonstrated in \asp\ via an
M-theory lift. A crucial ingredient has been the proposal,
based on a consistenct argument,
that a particular $U(1)$ orbit of the gauged LSM 
serves as an M-theory fiber. Specifically, the D-term 
equations for the $U(1)$ gauge symmetry of the LSM are
\eqn\dterms{
\sum_i \lm \al_i\, |x_i|^2 = r_\al,\ \ \ \lm \al_i\in\ZZ.
}
The proposal says that the corresponding $U(1)$
orbit $$x_i \to e^{i\sum \lm \al _i \theta_i}x_i$$
serves as the fiber of the M-theory lift for the theory 
with flux through the 2-cycle dual to the charge vector $\lm a$. 
This ansatz may now be understood as a consequence of the 
world-volume analysis of ref.\AViv.

\subsec{D-branes on the mirror $Y^*$ and a gauged LSM for them}
The starting point will be the gauged LSM for the closed string
background, namely the CY 3-fold $Y^*$ mirror to $Y$. The manifold
$Y^*$ is the total space of the canonical bundle $K(S)$ 
over a compact toric variety $S$ with $c_1>0$.
For simplicity we will assume dim$_\CC(S)=2$ so that
there is one non-compact direction; the necessary modifications
for dim$_\CC(S)=1$ are straightforward.
The gauged LSM describes the 
\CY $Y^*$  as the vacuum of a
2d (2,2) supersymmetric gauge theory with gauge group $U(1)^N$,
$N+3=h^{1,1(Y^*)}+3$ matter multiplets $x_i$ with charges
$\lm a$ under the $a$-th $U(1)$ factor and zero superpotential \wlsm.
The vacuum is the space of solutions of the D-term equations
\dterms\ divided by the $U(1)^N$ gauge symmetry.
The constants $r_a$ are the FI parameters which combine with the
integrals of the B-fields into the complex scalar fields $t_a$.
Alternatively, $Y^*$ is the symplectic quotient
$$
\Xi\, \backslash
\CC^{N+3}/(\CC^*)^N,\qquad (\CC^*)^N:\ x_i\to\lambda^{\lm a_i}x_i,\
\ \lambda \in \CC^*,
$$
where $x_i$ are coordinates on $\CC^{N+3}$ and $\Xi$ is the
fixed point set of the $(\CC^*)^N$ action \wlsm\mlsm.
The toric data for this quotient are summarized in the toric polyhedron
$\Delta\in \ZZ^3$, spanned by the $N+3$ vertices $\nu_i$ that satisfy
the relation $\sum_i \lm a _i \nu_i = 0$.\foot{As there 
is no superpotential in the A-model, the only moduli are the K\"ahler
moduli of $Y^*$ and the equivalent data of the complex structure of $Y$.
Both spaces are encoded in the single polyhedron $\Delta$.}

The D-branes on sL 3-cycles on $Y^*$ may be defined classically by
the equations \AVi
$$
\sum_i q_i^\al|x_i|^2=c_\al, \qquad
\theta^i=q_i^\al\phi_\al,\qquad   \al=1,\dots,3-k,
$$
where $x_i=|x_i|e^{i\theta_i}$, the $c_\al$ are 
complex constants and $q_i^\al\in \ZZ$ are integers that
satisfy  $\sum_i q_i^\al=0$. More precisely, $L$
is a complete 3-manifold if it ends on an edge
$|x_i|^2-|x_j|^2=0$ of the image of the moment map $x_i\to |x_i|^2$;
otherwise one has to add another component \AVi. 
For $k=2$ the topology of $L$
is $\bx R_{\geq 0}\times S^1 \times S^1$. Mirror symmetry on the
sL $T^3$ fiber defined by the independent phases of the $x_i$ removes
the two $S^1$'s and adds a different $S^1$, resulting in a
non-compact 2-cycle $C$ of topology $S^1\times \bx R_{\geq0}$
as in \defc.

More explicitly, mirror symmetry predicts the relation
$|y_i|=e^{-|x_i|^2}$ \HV, and a sL cycle $L$ mirror to the
2-cycle $C$ \defc\ is defined by the equations 
\eqn\dbrane{
|x_1|^2-|x_3|^2=c_1,\qquad
|x_2|^2-|x_3|^2=c_2,\qquad
\sum_i\theta_i=\rmx{const}.
}
Moreover the open string modulus $z_0$ in \defbound\ is related
to the above constants by \eqn\osma{|\zo| = e^{-c_1},}
as follows from \defboundii.
For an appropriate choice 
of coordinate labels, these equations describe D-branes of the
chosen topology in any patch of the CY manifold. It is  
assumed that this choice has been made such that $e^{-c_1}\sim |\zo|\leq1$.

The toric geometry that encodes the LSM for $Y^*$ {\it and} the
above D-branes on it is defined by a polyhedron $\Delop$ 
obtained from the polyhedron $\Delta$ for $Y^*$ as follows. 
The new polyhedron $\Delop\in \ZZ^4$ contains
the vertices of the polyhedron $\Delta$ on a hyperplane $H$, 
say $\nu'_i=(\nu_i,0),\, i=1,...,N+3$.
In addition $\Delop$ has two extra vertices 
$\nu'_{N+4}=(\nu_i,1)$ and $\nu'_{N+5}=(\nu_j,1)$ above $H$:
\eqn\convexhull{
\Delop = \rmx{convex\
hull}\pmatrix{(\nu_1,0)\cr\vdots\cr(\nu_{N+3},0)\cr
(\nu_i,1)\cr(\nu_j,1)\cr}.
}
As before, the integers $(i,j)$ 
may be chosen general, and $\Delop$ describes the mirror of the D-brane 
on \defc\ in a phase parametrized as in \tdspii\ for $(i,j)=(1,3)$. 
In the above we have used the standard correspondence of toric geometry 
that assigns a specific vertex $\nu_i$ to the
coordinate $x_i$ (and a monomial $\ty_i$ in the mirror $Y$) \mlsm.
{\goodbreak\midinsert
\centerline{\epsfxsize 3.1truein \epsfbox{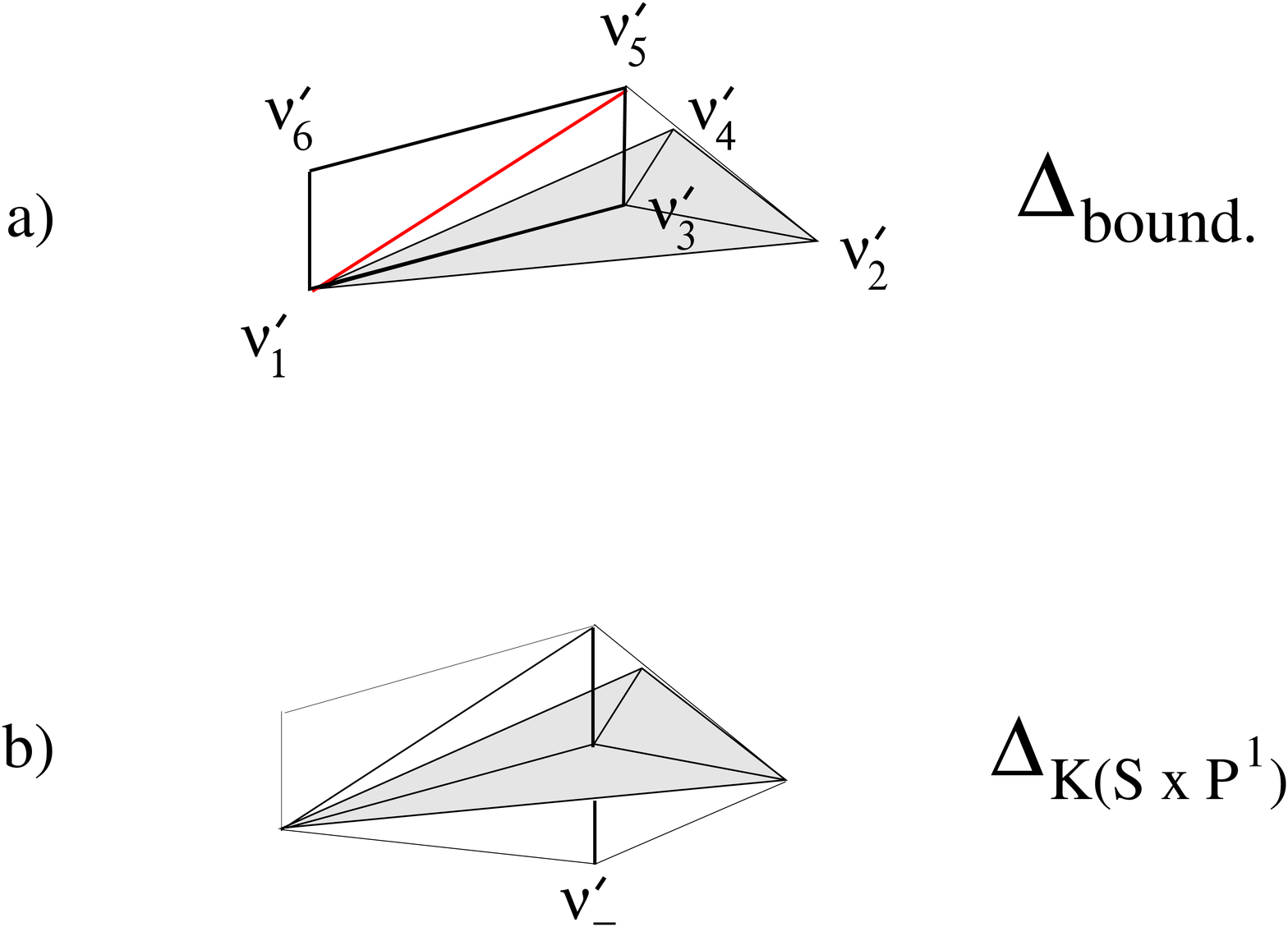}}
\leftskip 1pc\rightskip 1pc
\noindent{\ninepoint
{\bf Fig. 1}: $a)$ The toric polyhedron $\Delop$ that defines
the LSM for the CY $Y=K(\IP^2)$ with D-branes; $b)$ The enlarged
polyhedron $\Delta'$ that describes a fibration with base $\IP^1$.
}\endinsert}\vskip -0cm\ni
In LSM terms we have added two new matter fields uncharged
under the $U(1)^N$ gauge group and one new $U(1)$ gauge symmetry
that acts as 
$$
(x_i,x_j,x_{N+4},x_{N+5}) \ \to \ 
(\lambda\, x_i,\lambda^{-1}\, x_j,\lambda\, x_{N+4},\lambda^{-1}\, 
x_{N+5})\, ,
$$
with the other fields invariant.
To understand better the geometry defined by $\Delop$ we may
add a further vertex $\nu'_-$ below the hyperplane $H$ to obtain
an extended polyhedron $\Delta' \in \ZZ^4$. The toric geometry
$X'$ associated to $\Delta'$ is the canonical bundle of a 
fibration of $S$ over $\IP^1$, with the position 
of the point $\nu'_-$ specifying the fibration. Moreover one
of the two extra vertices above $H$ represents a blow
up of this fibration above a point $p$ on the base $\IP^1$.
Removing $\nu'_-$ again corresponds to taking the large 
volume limit of the base $\IP^1$ of the fibration 
and concentrating on a neighborhood of the special point $p$ 
isomorphic to $\CC$.

In fact geometries of this type have already played a role
in the context of F-theory and Abel-Jacobi maps in \bkmv.
For instance, for $S=\IP^2$ and choosing a non-trivial fibration,
the hypersurface $f=0$ in the toric variety $X'$ defined by $\Delta'$
is an elliptically fibered K3
surface with Picard lattice of rank three.
The two points $\nu'_k,\, k>N+3$ correspond to the
blow up of a singular fiber of ``$A_1$ type'' in the
elliptic fibration\foot{Specifically, the resolution replaces
the singular fiber by two spheres intersecting according to the
affine Dynkin diagram of $A_1$.}. Finally removing $\nu'_-$ corresponds
to the large base limit and concentrating on a neighborhood
of the $A_1$ fiber in the fibration. The moduli of this local
manifold is the same as that of a flat $SU(2)$ bundle on the
torus \bkmv, which in turn is the same as a point on the dual
torus in which the Abel-Jacobi map takes value.
In fact such a relation holds for any simply laced group $G$
instead of $A_1$, which may be realized \cf\ by rank($G)+1$
vertices above the hyperplane $H$.

A similar comment applies also for other choices for $S$. Note that
the interpretation of the polyhedron $\Delop$ as defining the
ambient space for an embedded K3 surface implies that the solutions 
of the GKZ system \defGKZii, which describe the open string superpotential,
are in fact related to the periods of K3 manifolds. This
follows from the relation \HIV\
\eqn\noncomprel{
\Pi_{comp.}=\fc{\p}{\p t}\Pi_{non-comp.}
}
between the periods of $Y^*$ and those of 
a compact K3 manifold embedded in it. It would be
interesting to study the implications of this further.

\subsec{K\"ahler cones, (D-brane) flops and existence of vacua}
We will now use the LSM to study the K\"ahler cones of the $\cx N=1$
moduli space and phase transitions between them. The moduli space 
of the gauged LSM will have in general several
limit points of maximally unipotent monodromy which, in the present
context, correspond to the classical vacua of the D-brane geometry.
More precisely the superpotential at these points 
is entirely non-perturbative.
This is reflected in the relation $z_\al\sim e^{2\pi i t_\al}$ 
and the polynomial dependence \Wouter\ of the superpotential $W(z_\al)$ 
near these limit points.

Different limit points are distinguished by different choices for a 
basis $\lm \al$ of the integral charge lattice. 
The correct basis is distinguished by
the property that the moduli $t_\al$ measure the sizes of the
fundamental holomorphic world-sheet maps of minimal volume.
That is, the minimal volume of a holomorphic world-sheet in any 
homology class must be a {\it positive} linear combination of the $t_a$. 
The existing limit points and the associated basis for 
the $\lm \al$ may be determined by studying the triangulations of the 
toric polyhedron $\Delop$ \mlsm.
A change between different bases amounts to taking
integral linear combinations
$\lm \al \to M_{\al\be}\lm \be$ and is accompanied by a 
redefinition of the flat coordinates $t_\al \to M_{\al\be}t_\be$. 

In general there are two, physically very different types of 
classical phases for the D-branes on $L$ in \dbrane, which are
reflected in two different types of triangulations of $\Delop$.
In the first case, referred to as an {\it inner phase} in the following, the
D-brane ends on the compact zero section $S:\, x_0=0$ of the \CY $Y^*$.
Note that the closed string degrees of freedom coming with $Y^*$ are 
localized on $S$ and thus the position of $L$ relative to $S$ 
determines the open-closed string interactions.
This inner phase corresponds to a so-called star triangulation of $\Delop$,
where all vertices are connected to the distinguished vertex $\nu_0$.
In contrast, an {\it outer phase} is defined by the D-brane ending
on a vertex $x_i=x_j=0,\ i,j\neq 0$ away from the zero section.
In this case one of the extra vertices representing the
D-brane does not lie in a cone together with $\nu_0$.

As will be argued below, the different open-closed string interactions 
in these two types of phases lead to quite different vacuum structures.
Whereas the D-brane in the outer phase 
may be treated as a perturbation of the CY background and has always a 
vacuum solution, the nearly classical D-brane in the inner phase 
destabilizes the CY geometry $Y^*$. 

To avoid inflated notation, we will describe 
some general features of these limit points,
the associated triangulations and
how they encode the physics of the superpotential mainly
at the hand of a simple example. The adaptation
to other situations is straightforward; see also
Appendix A for other examples. We consider 
again the non-compact \CY $Y^*=K(\IP^2)=\cx O(-3)_{\IP^2}$. 
The vertices for the toric polyhedron $\Delta$ and
the single charge vector $\lm1$ are
\eqn\Ypoly{
\eqalign{
\nu_0&=(0,0),\ \nu_1=(1,1),\ \nu_2=(-1,0),\ \nu_3=(0,-1),\cr
&\hskip50pt \lm 1= (-3,1,1,1),
}}
where here and in the following we distinguish the 
non-compact direction of $Y^*$ by the label ``$0$'' 
as in \mori.
The LSM for the D-branes on $Y^*$ will be defined by the toric polyhedron
$\Delop$ with vertices 
\eqn\polyi{\eqalign{
\nu_0&=(0,0,0),\ \nu_1=(1,1,0),\ \nu_2=(-1,0,0),\cr
\nu_3&=(0,-1,0),\ \nu_4=(1,1,1),\ \nu_5=(0,0,1).
}}

\ni
As alluded to above, the D-brane in an outer phase
may be treated as a perturbation, and as a consequence
the appropriate coordinates on the open-closed string moduli 
space should agree with the coordinates on the closed string moduli space, 
plus an extra modulus for the D-brane. This is confirmed by the existence of
a non-star triangulation of $\Delop$ leading to a basis of charge vectors 
of the form \mori
$$
{\eqalign{
\lm a&=(\lm a(Y^*)\hskip20pt ;0,\ \, 0),\qquad a=1,...,h^{1,1}(Y^*),\cr
\lm 0&=(1,-1,0,...0;1,-1).
}}
$$
Here the first entry corresponds to the non-compact direction
and the last two entries to the new vertices not contained in the
hyperplane $\nu_{i,3}=0$. 
The point of maximal unipotent monodromy is specified by $z_\al=0$.
This translates by \osma\ to the condition $|x_0|^2\gg |x_1|^2$, which indeed
identifies this triangulation as the outer phase. 
This phase is sketched on the l.h.s. of Fig.2. 
In this non-star triangulation, 
the point $\nu'_6$ is not connected to the distinguished vertex $\nu_0$ 
and can be decoupled (decompactified). 
Physicswise the semi-classical D-brane does not destabilize 
the CY geometry $Y^*$, and has a runaway vacuum where the D-brane moves
to an infinite distance away from the compact divisor $x_0$ that supports
the closed string excitations. 
{\goodbreak\midinsert
\centerline{\epsfxsize 4.3truein \epsfbox{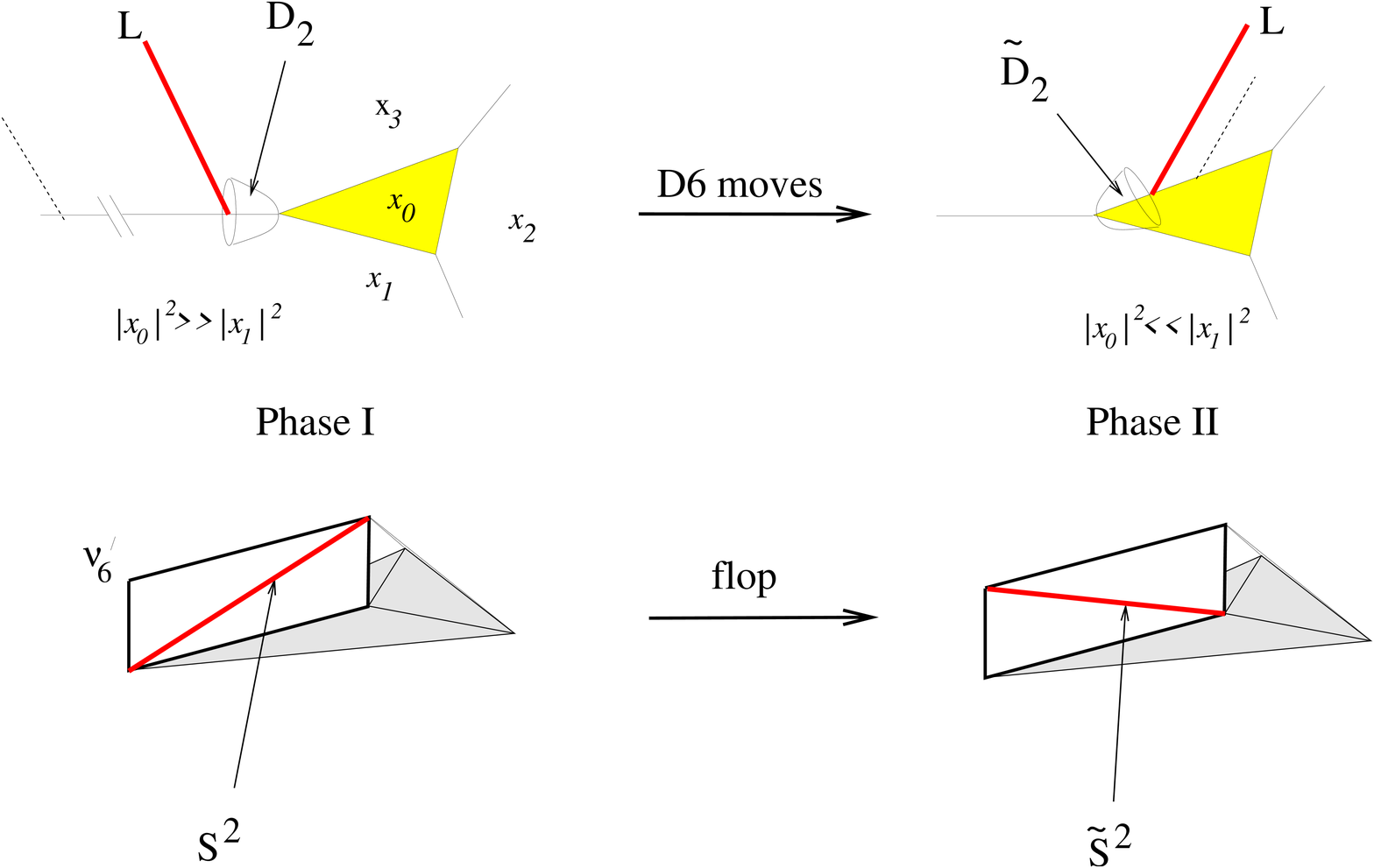}}
\leftskip 1pc\rightskip 1pc 
\noindent{\ninepoint  
{\bf Fig.~2}: The (D-brane) flop transition in the 
open string picture and in the associated LSM defined by $\Delop$.
In the former, the D-brane on $L$ moves (in the image of the moment map 
$x_i\to |x_i|^2$) towards a vertex of $\IP^2$, where the area of
a disc instanton on $D_2$ vanishes, and continues to the 
other side where a new disc instanton appears on $\tx D_2$. 
In the LSM the same phase transition is described by the 
flop in the (image of the) triangulation for $\Delop$.
}\endinsert}

\ni
For increasing $\zo$, the perturbative expansion \Wouter\ 
fails to converge near the wall Re$(\zo)\sim 1$ or $t_0 \sim 0$, 
which is a highly non-classical regime from the point 
of string world-sheet expansion. Continuing further to 
large $\zo$ one finds a new point of maximal unipotent
monodromy corresponding to a new classical limit. In the 
classical D-brane picture, the end of the brane on $x_i=x_j=0$
has moved to an inner vertex $x_0=x_i=0$ which intersects
the zero section $S$. In the LSM this ``transition'' amounts to
a flop in the triangulation, as illustrated in Fig.~2\foot{Note however
that the classical definition of the D-brane after the flop
involves also a shift of the zero energy. This is related to the 
change of the classical terms and their subtraction in the A-model
(or a change of the reference curve $C_*$ in \chainsp\ in the B-model).
The associated classical limits are indicated as dashed lines in
Fig.2.}.
A basis of charge vectors associated to the new triangulation is
\def\ss{}
{\vbox{{
\vskip 0.1cm
\eqn\morii{
\vbox{\offinterlineskip\tabskip=0pt\halign{\strut
\hfil~$\ss{#}$~\hfil
&\hfil$\ss{#}$~
&\hfil$\ss{#}$~
&\hfil$\ss{#}$~
&\hfil$\ss{#}$~
&\hfil$\ss{#}$~
&\hfil$\ss{#}$~
&\hfil$\ss{#}$~
&\hfil\hfil$\ss{#}$~
&$\ss{#}$~\cr
\tlm1\,=\,(&-2& 0& 1& 1&-1& 1&)&,  \cr
\tlm0\,=\,(&-1& 1& 0& 0& 1&-1&)&,  \cr
}}}
\vskip -0.25cm}}}
\ni
and leads by \defz\ to new coordinates in a patch $\tx z_\al=0$.

Note that this choice of coordinates mixes the 
notion of closed and open string moduli, in contrast to
the classical picture where the D-brane just perturbs
the CY moduli space. In particular the non-trivial
mixing has the effect that the superpotential 
in this phase destabilizes the CY geometry near the new
expansion point $\tx z_\al=0$. 
Specifically, analogous to what we did in section 2.3, we can determine
the expansion of the superpotential in the inner phase to be: 
\eqn\Winner{
W(z_\a,z_0) = \!\!\!
\sum_{{n_\a\geq0,n_0\geq 0, n_0\not=n_1
\atop n_0-\sum\tilde\lia\a_0n_\a>0}}
\!\!
{(-)^{-n_1+\sum\tilde\lia\a_0n_\a}(n_0-\sum\tilde\lia\a_0n_\a-1)!
\over
(n_0\!-\!n_1) (n_0\!+\!\sum\tilde\lia\a_1n_\a)!
\prod_{i=2}^{N+2}(\sum\tilde\lia\a_in_\a)!}{z_0}^{n_0}
\!\!\prod_{\a}{z_\a}^{n_\a}.
}
This expression is actually more general\foot{With some little more effort 
it is possible to write
down a superpotential for any given kind of flop.} 
and applies to all 
flops in CY geometries with charge vectors \mori\
for which the new charge vectors can be represented by
$\tlm 1 = \lm 1 +\lm 0$, $\tlm 0=-\lm 0$.

It follows straightforwardly from the expression above that,
different as compared to the superpotential in the outer phase, 
the superpotential \Winner\ has no solution $dW=0$ at finite 
volume of the \CY $Y^*$ near the limit point $z_\a=0$.
This behavior is also in agreement with the 
classical picture that the volumes of the discs in this phase
are bounded by the volume of 2-spheres and thus the
disc instantons drive the CY towards its decompactification 
(along its compact direction)%
\foot{The structure of 
the superpotential in the inner and outer phases parallels
that of  $\cx N=1$ SYM theories 
with few matter multiplets embedded in string theory,
with masses below and above the field theory scale $\Lambda$, 
respectively.
In view of the dualities of \vafaln\ there may well be a 
concrete correspondence of this kind in a certain
regime of the moduli space.}.

\newsec{Discussion}
As argued in \asp, the existence of a Picard-Fuchs system for the $\cx N=1$
superpotential of the open-closed string system reflects a new
structure, namely the $\cx N=1$ special geometry of the deformation space 
of the topologically twisted theories. This
is explicit in the open/closed duals studied there, where the
$\cx N=1$ special geometry derives from the special
geometry of CY 4-folds. We have not yet discussed the 
non-holomorphic content of this geometry of the $\cx N=1$ ``moduli space'',
which is governed by the $tt^*$ equations of \refs{\tts,\bcov}.
In particular we have not considered
the kinetic terms of the chiral multiplets
for the ``moduli'' fields. This is  an interesting 
subject for further study. 

Here we restrict ourselves to point out
one particularly interesting aspect of the kinetic terms, namely the relation
to the theory of modular functions for the CY moduli space
and open string generalizations thereof. The holomorphic
superpotential $W$ enters the effective $\cx N=1$ four-dimensional
supergravity theory  via the function
$$
G=K(\phi_i,\bb \phi_i)+\ln W(\phi_i)+\ln \bb W (\bb \phi_i),
$$
where $K$ is the K\"ahler potential. $\cx N=1$ special 
geometry predicts a relation between $K$ and $W$.
On the other hand, in general supergravity theories the two
functions $K$ and $W$ are completely independent. 
However an $\cx N=1$ string effective supergravity from
the type II string on a \CY manifold will have quite
generally discrete quantum symmetries 
which imply a strong correlation between the K\"ahler 
and the superpotential. In fact the assumption of the
invariance of the function $G$ under a given modular group $SL(2,\ZZ)$ 
has been used in \ft\ to predict exact instanton corrected 
formulae for the K\"ahler potential and the superpotential 
starting from the classical expressions. 
The same line of thought should be quite powerful also in the open-closed
string systems considered in this paper, where the exact 
expression for $W$ is known and might be used to 
determine the kinetic terms. The open string superpotential 
contributes the term
$$
\ln W=\ln\big(\sum_{k=1}^{\infty} q_0^k\, f_k(q_a)\big)=
2\pi i t_0 +F(q_a) +\sum_{k=1}^\infty q_0^k\, \tx f_k(q_a)
$$
to the $G$-function. The functions of the closed string moduli $q_a$ 
in the above expression have modular properties 
with respect to the modular
group $\Gamma$ of the complex structure moduli space of the CY 3-fold. 
E.g. the leading term for large vev's of the open string modulus
$t_0$  comprises the instanton corrections \linco\ to the open string 
mirror map and is of the form 
$$F(q_a)=\sum_a \fc{b_a}{2\pi i}\ln(q_a/z_a(q_a)).$$ Here the $b_a$ are some
rational coefficients determined in \linco\ and the non-trivial
functions $z_a(q_a)$ have a relatively 
straightforward interpretation in terms of modular functions of $\Gamma$.
The higher orders in $q_0$ should describe the embedding of 
$\Gamma$ into the larger modular group $\Gamma'$ of the
total open-closed string moduli space\foot{Again it is helpful 
to think of the group $\Gamma'$ in terms of the 
 modular group of the complex structure
moduli space of a CY 4-fold via the open/closed string dualities
in \asp.}. This is similar to the embedding of modular groups of
K3 into modular groups of CY 3-folds studied in \modfcts.
It would be interesting to study the quantum symmetries
and their implications for the metric and extrema of the superpotential
further.

Another comment concerns the fact that in the approach of \AVi,
there is an integral ambiguity in the non-perturbative
definition of the open-closed string system \AVii. This is related to 
the framing ambiguity of knots in the
Chern-Simons theory on the internal part of the D-brane.
On the other hand the definition of the mirror map by the
Picard-Fuchs system for the period and chain integrals 
does not have such an ambiguity and thus
provides a preferred framing. Different framings correspond to 
different parametrizations of the curve $F=0$ with $F$
defined in \tdspii. It is straightforward to see that the preferred framings
are related to those parametrizations of $F$ that are obtained from
the 2d superpotential $\Wtd$ by factoring out a coordinate $\ty_i$
{\it linearly}. This is in fact a necessary condition in order
that the two superpotentials \mirrorsp\ and \tdspii\ are
equivalent on the level of period integrals \HIV, and thus
explains the selection of framing made by the PF system.

\vskip 1cm
\ni \bx{Acknowledgments:} We are grateful to
I. Brunner and
S. Stieberger
for discussions.
\vfil\goodbreak

\appendix{A}{Picard-Fuchs equations, linear sigma models and 
superpotentials from toric geometry}

We present below a collection of toric computations for D-branes on
non-compact 3-folds, given by the canonical bundles over $\IP^2$ and
the Hirzebruch surfaces $F_0$, $F_1$ and $F_2$.
For those cases that have been discussed already in
refs.~\refs{\AVi,\AVii}\foot{See also \IK\ for similar
computations for other CY geometries.},
our results are consistent with the known results\foot{%
With two exceptions: the instanton numbers given in \AVii\ for
the outer phase of $F_0$ and an inner phase of $F_1$; see below.}.
The closed formulas that we obtain allow 
to determine the disc instanton numbers
$N_{n_\a,n_0}$ for all $n_0$ explicitly, 
for given definite values of $n_\a$. This is also the natural form of 
the results obtained by the
localization techniques of ref.\zas.
{From} the point of view of differential equations,
their origin may be traced back to 
recursion relations for the instanton numbers derived from the
boundary differential operator $\cx D_0$, which are quadratic in $n_0$
and determine $N_{n_a,n_0}$ in terms of the $N_{n_a,n_0-1}$.
We will consider two phases for each case,
the first corresponding to an outer phase and the
second to an inner phase obtained by a flop, as discussed in 
section~3.

\goodbreak
\subsec{\someline$Y^*=K(\IP^2)$\someline}

The application of the GKZ system \defGKZii\ to this case
has been already sketched in \asp\ and we restrict ourselves below to a 
complete the discussion and to add some results.
The polyhedron $\Delop$ and a basis of charge vectors for the outer phase 
have been defined in \polyi\ and \mori. The associated Picard-Fuchs system 
is 
$$
\eqalign{
\cD_1&=
{{\xt(1)}^2}\left( \xt(1) - \xt(0) \right)  +
  \xz(1)\left( 3\xt(1) - \xt(0) \right)
   \left( 1 + 3\xt(1) - \xt(0) \right)
   \left( 2 + 3\xt(1) - \xt(0) \right)\ ,
\cr
\cD_0&=
\left( 3\xt(1) - \xt(0) \right) \xt(0)-
\xz(0)\left( \xt(1) - \xt(0) \right) \xt(0)
\ .}\weqn\PFPtwoIII
$$
The corrections to the mirror map \flatco\ are  $S_0=A$, $S_1=-3\, A$,
with 
$$
A=-\sum_{{n_1>0}}
{(-)^{n_1}(3n_1-1)!\over {n_1!}^3}
 {z_{1}}^{n_1}
\ .\weqn\APtwoIII
$$
The superpotential \Wouter\ is given by
$$
W(z_1,z_0)=\sum_{{n_1\geq0,n_0>n_1}}
{\frac{{{\left( - \right) }^{\xa(1)}}
     \left( \xa(0) - \xa(1) -1\right) !}{\left( \xa(0) -
        3\xa(1) \right) !{{\xa(1)!}^2}\xa(0)}}
{z_0}^{n_0} {z_1}^{n_1}
\ .\weqn\SPPtwoIII
$$
The above expression 
can be rewritten in terms of hyper-geometric functions as in 
\Whyp\ and \hypid, 
and moreover taking a derivative and using \ftwosimp\ we 
have
$$
\eqalign{
z_{0}\del_{z_0}W(z_1,z_0) 
&=\sum_{n\geq0}
{(-)^n{z_0}^{n+1}{z_1}^{n}\over {n!}^2\Gamma(2-2n) }
\,{}_2F_1(1,1,2-2n;z_0)
\cr&=
-\log(1-z_0)+\sum_{n\geq0}
{(-)^n(2n-1)!\over  {n!}^2 }
\left(z_0\over 1-z_0\right)^{2n}\!\!(z_0z_1)^n
\cr&=
-\log
\left(
\Coeff12\left[1-z_0+\sqrt{4{z_0}^3z_1+(z_0-1)^2}\right]
\right)\ .
}
$$
The last expression is the logarithm of the solution for $z_u$ 
of the equation $z_u+z_0-z_1{z_0}^3{z_u}^{-1}-1=0$, 
which reproduces the Riemann surface of ref.\ \AVii.
Note that the expression in the square root is a component of
the discriminant
of the PF system \PFPtwoIII; this appears to be a general feature.

Our explicit formula allow to extract closed expressions
for the disc instanton numbers. Concretely we find for
the first generating functions $N_{n_1}(x)= \sum_{n_0} N_{n_1,n_0}
x^{n_0}$: 
\def\coeff#1#2{\relax{\ss {#1 \over #2}}}
\def\ss{\scriptstyle}
$$
\vbox{
\offinterlineskip\tabskip=0pt\halign{\strut
\vrule~$\ss #$~&\vrule\vrule~$\ss #$~\vrule\cr
\noalign{\hrule}
1& {\frc{x\left(  2-x \right)}{x-1}}\cr
\noalign{\hrule}
2& {\frc{x\left( -5 + 6x + 5{x^2} - 8{x^3} + {x^5} \right) }
   {{{\left( x-1 \right) }^3}\left( 1 + x \right) }}\cr
\noalign{\hrule}
3& \frc{x}{{{\left( 1-x \right) }^5}\left( 1 + x + {x^2} \right) }
( -32 + 107x - 126{x^2} + 86{x^3} - 109{x^4} +
        125{x^5}
- 56{x^6} + 2{x^7} + {x^9} )
\cr
\noalign{\hrule}
}}\hfil
\weqn\ptwoNxIII
$$
These are consistent with the results listed in Table 6 of ref.\AVii\
obtained by the method of \AVi.
Closed formulae for the invariants $N_{n_1,n_0}$ valid for $n_0>n_1+1$ 
are:\foot{$\epsilon(n,x)$ is defined to be equal to be one if
$x\, ({\rm mod}\ n)=0$, and zero otherwise. The dependence of our expressions
on $\epsilon(n,x)$, with a coefficient proportional to the inverse square,
should reflect the open string instanton 
multi-covering formulae of ref.\ \vm.}  
$$
\eqalign{
N_{1,n_0}&=-1
\cr
N_{2,n_0}&=
{\Frcb{15 - 4\xa(0) + {{\xa(0)}^2}}{4}} + \coeff14\epsilon (2, \xa(0))
\cr
N_{3,n_0}&=
{\Frcb{-976 + 348\xa(0) - 103{{\xa(0)}^2} + 12{{\xa(0)}^3} - 
     {{\xa(0)}^4}}{36}} + \coeff19\epsilon (3, \xa(0))
\cr
N_{4,n_0}&=
\coefff{1}{576}(147996 - 59400\xa(0) + 18961{{\xa(0)}^2} - 
     3072{{\xa(0)}^3} + 370{{\xa(0)}^4}
\cr&\qquad
 - 24{{\xa(0)}^5} + 
     {{\xa(0)}^6}) - \coefff1{64}\left( 60 - 8\xa(0) + {{\xa(0)}^2} \right) 
      \epsilon (2, \xa(0))
\cr
}\weqn\zaspolptwoIII
$$
with $N_{n_1,0}\equiv0$.

\goodbreak

The LSM charge vectors in the flopped phase have been given in \morii.
The associated Picard-Fuchs operators are:
$$
\eqalign{
\cD_1&=
{{\xt(1)}^2}\left( \xt(1) - \xt(0) \right)  +
  \xz(1)\left( \xt(1) - \xt(0) \right)
   \left( 2\xt(1) + \xt(0) \right)
   \left( 1 + 2\xt(1) + \xt(0) \right)
\cr
\cD_0&=
\xt(0)\left( -\xt(1) + \xt(0) \right)  +
  \xz(0)\left( \xt(1) - \xt(0) \right)
   \left( 2\xt(1) + \xt(0) \right)
\ .}\weqn\PFPtwoI
$$
The corrections to the mirror map are $S_0=-A$, $S_1=-2\, A$,  with 
$$
A(z_1,z_0)=-\sum_{{n>0}}
{(-)^{n}(3n-1)!\over {n!}^3}
 (z_0z_{1})^{n}
\ .\weqn\APtwoI
$$
The superpotential \Winner\ becomes:
$$
W(z_1,z_0)=\sum_{{n_{0,1}\geq0,n_0\not=n_1}}
{\frac{{{\left( - \right) }^{\xa(1)}}\left( \xa(0) + 2\xa(1) -1\right)
!}
   {\xa(0)!{{\xa(1)!}^2}
     \left( \xa(0) - \xa(1) \right) }}
{z_0}^{n_0} {z_1}^{n_1}
\ .\weqn\SPPtwoI
$$

\ni The first generating functions $N_{n_1}(x)= \sum_{n_0} N_{n_1,n_0}
x^{n_0}$ \Nx\ for instanton numbers are 
\def\frcb#1#2{{\coeff1{#2}(#1)}}
\def\ss{\scriptstyle}
$$
\vbox{
\offinterlineskip\tabskip=0pt\halign{\strut
\vrule~$\ss #$~&\vrule\vrule~$\ss #$~\vrule\cr
\noalign{\hrule}
1& {\frcb{x-1 + {x^2}}{x-1}}\cr
\noalign{\hrule}
2& {\frcb{1 - 4{x^2} - 3{x^3} + 6{x^4} + 3{x^5} - 4{x^6}}
   {{{\left( x-1 \right) }^3}\left( 1 + x \right) }}\cr
\noalign{\hrule}
3&\frc{(-1 - 2{x^2} + 29{x^3} - 17{x^4} - 53{x^5} + 49{x^6} -
     9{x^7}}{{{\left(
          x-1 \right) }^5}\left( 1 + x + {x^2} \right) }
+ 55{x^8} - 76{x^9} + 27{x^{10}}) \cr
\noalign{\hrule}
}}\hfil\weqn\ptwoNxI
$$
Closed formulae for some invariants $N_{n_1,n_0}$ valid for $n_0>n_1+1$ are:
$$
\eqalign{
N_{1,n_0}&=-1
\cr
N_{2,n_0}&=
{\Frcb{11 + {{\xa(0)}^2}}{4}} + \coefff14\epsilon (2,\xa(0))
\cr
N_{3,n_0}&=
{\Frcb{-616 + 54\xa(0) - 49{{\xa(0)}^2} - {{\xa(0)}^4}}{36}} + 
     \coefff19\epsilon (3,\xa(0))
\cr
N_{4,n_0}&=
\coefff1{576}(91404 - 14976\xa(0) + 6097{{\xa(0)}^2} - 288{{\xa(0)}^3} + 
     130{{\xa(0)}^4} + {{\xa(0)}^6})
\cr&\qquad  - 
     \coefff1{64}\left( 44 + {{\xa(0)}^2} \right) \epsilon (2, \xa(0))
\cr
}\weqn\zaspolptwoI
$$
with $N_{n,n}\equiv0$. These expressions are again consistent with 
Table 5 of ref.\AVii\ after the shift of labels $n_0 \to n_0 - n_1$. 
See \asp\ for an explanation of the origin of the shift and Table 3
therein for an explicit listing of the above
disc numbers.

\goodbreak
\subsec{\someline$Y^*=K(F_0)$\someline}
The LSM for D-branes on the canonical bundle of $F_0$ is
defined by the toric polyhedron $\Delop$ with vertices
$$
\eqalign{
\nu_0&=(0,0,0),\ \nu_1=(-1,0,0),\ \nu_2=(1,0,0),\ \nu_3=(0,-1,0),\cr
\nu_4&=(0,1,0),\ \nu_5=(0,-1,1),\ \nu_6=(0,0,1).
}
$$
In the outer phase the D-brane ends on the 
edge $|x_1|^2=|x_3|^2=0$ and 
the charge vectors are
$$
\matrix{\lm 1\cr\lm 2\cr\lm 0}\  = \
\matrix{
(&-2 & 1 & 1 & 0 & 0 & 0 & 0 &)\cr
(&-2 & 0 & 0 & 1 & 1 & 0 & 0 &)\cr 
(& 1 & 0 & 0 & -1 & 0 & 1 &-1&) \cr  }
\ .\weqn\CVFnullIII
$$
The period and chain integrals satisfy the generalized PF system
$$
\eqalign{
\cD_1&=
{{\xt(1)}^2} - \xz(1)\left( 2\xt(1) + 2\xt(2) - \xt(0) \right)
   \left( 1 + 2\xt(1) + 2\xt(2) - \xt(0) \right)
\cr
\cD_2&=
\xt(2)\left( \xt(2) - \xt(0) \right)  -
  \xz(2)\left( 2\xt(1) + 2\xt(2) - \xt(0) \right)
   \left( 1 + 2\xt(1) + 2\xt(2) - \xt(0) \right)
\cr
\cD_0&=
  \left( 2\xt(1) + 2\xt(2) - \xt(0) \right) \xt(0)
-\xz(0)\left( \xt(2) - \xt(0) \right) \xt(0)
\ .}\weqn\PFFnullIII
$$
The corrections to the logarithmic solutions that define the open string 
mirror map are $S_0= A$, $S_1=-2\, A $, $S_2=-2\, A $,
with
$$
A=-\sum_{{n_1,n_2\geq0\atop(n_1,n_2)\not=(0,0)}}
{(2n_1+2n_2-1)!\over {n_1!}^2{n_2!}^2}
 {z_{1}}^{n_1}{z_{2}}^{n_2}
\ .\weqn\AFnullIII
$$
The superpotential $W$ is given by the series expansion
$$
W(z_\a,z_0)=\sum_{{n_i\geq0,n_0>n_2}}
{\frac{{{\left( - \right) }^{\xa(2)}}
     \left(\xa(0) - \xa(2)-1 \right) !}{{{\xa(1)!}^2}
     \left( \xa(0) - 2\xa(1) - 2\xa(2) \right) !\xa(2)!\xa(0)}}
{z_0}^{n_0} {z_1}^{n_1} {z_2}^{n_2}
\ .\weqn\SPFnullIII
$$
The first generating functions $N_{n_1,n_2}(x)= \sum_{n_0} N_{n_1,n_2,n_0} x^{n_0}$
for the disc instantons are:
$$
\vbox{
\offinterlineskip\tabskip=0pt\halign{\strut
\vrule~$\ss #$~&\vrule\vrule~$\ss #$~&\vrule~$\ss #$~&\vrule~$\ss #$~\vrule\cr
\noalign{\hrule}
&0&1&2\cr
\noalign{\hrule}\noalign{\hrule}
0&0 & \ss x & \ss 0 \cr
\noalign{\hrule}
1
&{\frc{x}{1 - x}}
&{\frc{\left( x -3\right) x}{x-1}}
&{\frc{x\left( -5 + x + {x^2} \right) }{x-1}}\cr
\noalign{\hrule}
2
&{\frc{{x^3}}{{{\left( 1-x \right) }^3}\left( 1 + x \right) }}
&{\frc{x\left( -5 + 11x - 9{x^2} + {x^3} \right) }
   {{{\left( x-1 \right) }^3}}}
&{\frc{x\left( -35 + 46x + 23{x^2} - 52{x^3} + 6{x^4} +
       2{x^5} \right) }{{{\left( x-1 \right) }^3}
     \left( 1 + x \right) }}\cr
\noalign{\hrule}
}}\hfil\weqn\FnullNxIII
$$

The disc instanton numbers of low degee are collected
in the following table, with the horizontal (vertical) 
direction corresponding to $n_1$ ($n_2)$ and the bold
face letter in the corner denoting $n_0$:

\def\aa(#1){\hfil{\scriptstyle #1}}
$$
\vbox{
\offinterlineskip\tabskip=0pt\halign{\strut
$#$~\hfil\vrule&
\hfil~$#$~&\hfil~$#$~&\hfil~$#$~&\hfil~$#$~&
\hfil~$#$~&\hfil~$#$~&\hfil~$#$~&\hfil~$#$~&
\hfil~$#$~&\hfil~$#$~
\cr
{\bf 1}&0&1&2&3&4\cr
\noalign{\hrule}
0 &     \aa(1)&\aa( 1)&\aa( 0)&\aa( 0)&\aa( 0)\cr
1 &     \aa(1)&\aa( 3)&\aa( 5)&\aa( 7)&\aa( 9)\cr
2 &     \aa(0)&\aa( 5)&\aa( 35)&\aa( 135)&\aa( 385)\cr
3 &     \aa(0)&\aa( 7)&\aa( 135)&\aa( 1100)&\aa( 5772)\cr
4 &     \aa(0)&\aa( 9)&\aa( 385)&\aa( 5772)&\aa( 50250)\cr
5 &     \aa(0)&\aa( 11)&\aa( 910)&\aa( 22950)&\aa( 309638)\cr
6 &     \aa(0)&\aa( 13)&\aa( 1890)&\aa( 75174)&\aa( 1495832)\cr
}}\hskip25pt
\vbox{
\offinterlineskip\tabskip=0pt\halign{\strut
$#$~\hfil\vrule&
\hfil~$#$~&\hfil~$#$~&\hfil~$#$~&\hfil~$#$~&
\hfil~$#$~&\hfil~$#$~&\hfil~$#$~&\hfil~$#$~&
\hfil~$#$~&\hfil~$#$~
\cr
{\bf 2}&0&1&2&3&4\cr
\noalign{\hrule}
0&      \aa(0)&\aa( 0)&\aa( 0)&\aa( 0)&\aa( 0)\cr
1&      \aa(1)&\aa( 2)&\aa( 4)&\aa( 6)&\aa( 8)\cr
2&      \aa(0)&\aa( 4)&\aa( 24)&\aa( 96)&\aa( 280)\cr
3&      \aa(0)&\aa( 6)&\aa( 96)&\aa( 750)&\aa( 3936)\cr
4&      \aa(0)&\aa( 8)&\aa( 280)&\aa( 3936)&\aa( 33544)\cr
5&      \aa(0)&\aa( 10)&\aa( 684)&\aa( 15876)&\aa( 206656)\cr
6 &     \aa(0)&\aa( 12)&\aa( 1456)&\aa( 52992)&\aa( 1007208)\cr
}}\hfil
$$
$$
\vbox{
\offinterlineskip\tabskip=0pt\halign{\strut
$#$~\hfil\vrule&
\hfil~$#$~&\hfil~$#$~&\hfil~$#$~&\hfil~$#$~&
\hfil~$#$~&\hfil~$#$~&\hfil~$#$~&\hfil~$#$~&
\hfil~$#$~&\hfil~$#$~
\cr
{\bf 3}&0&1&2&3&4\cr
\noalign{\hrule}
0&      \aa(0)&\aa( 0)&\aa( 0)&\aa( 0)&\aa( 0)\cr
1&      \aa(1)&\aa( 2)&\aa( 3)&\aa( 5)&\aa( 7)\cr
2&      \aa(1)&\aa( 6)&\aa( 25)&\aa( 84)&\aa( 234)\cr
3&      \aa(0)&\aa( 10)&\aa( 112)&\aa( 729)&\aa( 3476)\cr
4&      \aa(0)&\aa( 14)&\aa( 360)&\aa( 4191)&\aa( 31876)\cr
5&      \aa(0)&\aa( 18)&\aa( 935)&\aa( 18187)&\aa( 210075)\cr
6&      \aa(0)&\aa( 22)&\aa( 2093)&\aa( 64395)&\aa( 1086215)\cr
}}\hskip25pt
\vbox{
\offinterlineskip\tabskip=0pt\halign{\strut
$#$~\hfil\vrule&
\hfil~$#$~&\hfil~$#$~&\hfil~$#$~&\hfil~$#$~&
\hfil~$#$~&\hfil~$#$~&\hfil~$#$~&\hfil~$#$~&
\hfil~$#$~&\hfil~$#$~
\cr
{\bf 4}&0&1&2&3&4\cr
\noalign{\hrule}
0&       \aa(0)&\aa( 0)&\aa( 0)&\aa( 0)&\aa( 0)\cr
1&        \aa(1)&\aa( 2)&\aa( 3)&\aa( 4)&\aa( 6)\cr
2&        \aa(2)&\aa( 10)&\aa( 32)&\aa( 90)&\aa( 224)\cr
3&        \aa(1)&\aa( 20)&\aa( 165)&\aa( 896)&\aa( 3775)\cr
4&        \aa(0)&\aa( 30)&\aa( 576)&\aa( 5650)&\aa( 38016)\cr
5&        \aa(0)&\aa( 40)&\aa( 1595)&\aa( 26316)&\aa( 269843)\cr
6&        \aa(0)&\aa( 50)&\aa( 3744)&\aa( 98588)&\aa( 1481984)\cr
}}\hfil
$$
$$
\vbox{
\offinterlineskip\tabskip=0pt\halign{\strut
$#$~\hfil\vrule&
\hfil~$#$~&\hfil~$#$~&\hfil~$#$~&\hfil~$#$~&
\hfil~$#$~&\hfil~$#$~&\hfil~$#$~&\hfil~$#$~&
\hfil~$#$~&\hfil~$#$~
\cr
{\bf 5}&0&1&2&3&4\cr
\noalign{\hrule}
0&      \aa(0)&\aa( 0)&\aa( 0)&\aa( 0)&\aa( 0)\cr
1&      \aa(1)&\aa( 2)&\aa( 3)&\aa( 4)&\aa( 5)\cr
2 &     \aa(4)&\aa( 16)&\aa( 45)&\aa( 110)&\aa( 245)\cr
3&      \aa(4)&\aa( 42)&\aa( 265)&\aa( 1232)&\aa( 4644)\cr
4 &     \aa(1)&\aa( 70)&\aa( 1015)&\aa( 8472)&\aa( 51018)\cr
5&      \aa(0)&\aa( 98)&\aa( 2997)&\aa( 42262)&\aa( 388063)\cr
6 &     \aa(0)&\aa( 126)&\aa( 7403)&\aa( 167440)&\aa( 2257665)\cr
}}\hskip25pt
\vbox{
\offinterlineskip\tabskip=0pt\halign{\strut
$#$~\hfil\vrule&
\hfil~$#$~&\hfil~$#$~&\hfil~$#$~&\hfil~$#$~&
\hfil~$#$~&\hfil~$#$~&\hfil~$#$~&\hfil~$#$~&
\hfil~$#$~&\hfil~$#$~
\cr
{\bf 6}&0&1&2&3&4\cr
\noalign{\hrule}
0&  \aa(0)&\aa( 0)&\aa( 0)&\aa( 0)&\aa( 0)\cr
1&      \aa(1)&\aa( 2)&\aa( 3)&\aa( 4)&\aa( 5)\cr
2&      \aa(6)&\aa( 24)&\aa( 62)&\aa( 140)&\aa( 288)\cr
3&      \aa(11)&\aa( 86)&\aa( 440)&\aa( 1782)&\aa( 6096)\cr
4&      \aa(6)&\aa( 168)&\aa( 1866)&\aa( 13400)&\aa( 72750)\cr
5&      \aa(1)&\aa( 252)&\aa( 5920)&\aa( 71680)&\aa( 592287)\cr
6&      \aa(0)&\aa( 336)&\aa( 15430)&\aa( 300672)&\aa( )\cr
}}\hfil
$$
\vskip10pt
\vbox{\leftskip 2pc\rightskip 2pc
\noindent{\ninepoint
{\bf Table A.1:} Disc instanton numbers for the D-brane ending 
on the outer edge $|x_1|^2=|x_3|^2=0$ of $F_0$.}}

\ni
These numbers disagree with those given in Table 3 of ref.\AVii,
computed in the approach of \AVi. A subsequent recalculation
of the superpotential with the methods of \AVi\ turns out to
be consistent with our result above.
Closed formulas for some invariants $N_{n_1,n_2,n_0}$ are
$$
\eqalign{
N_{1,n_2,n_0}& = n_2+1
\ ,n_0>n_2
\cr
N_{0,1,n_0}& = 0
\ ,n_0>2,\ \ N_{0,n_2,n_0} =0\ ,n_2>1,\forall n_0
\cr
N_{2,0,n_0}& =
{\Frc{{{\left(  \xa(0)-1 \right) }^2} }{4}}-\coefff14 \epsilon (2, \xa(0))
\ ,\forall n_0
\cr
N_{2,1,n_0}& =
6 - 3\,\xa(0) + {{\xa(0)}^2}
\ ,n_0>1
\cr
N_{2,2,n_0}& =
\coefff52\left( 13 - 4\,\xa(0) + {{\xa(0)}^2} \right)  - 
  {\coefff12{\epsilon (2,\xa(0))}}
\ ,n_0>2\ .
\cr
}\weqn\zaspolFnullIII
$$

\goodbreak
After the (flop) transition to the inner phase, the D-brane ends
on the egde $|x_1|^2=|x_0|^2=0$ and the basis of charge vectors becomes
$$
\matrix{\lm 1\cr\lm 2\cr\lm 0}\  = \
\matrix{
(&-2 & 1 & 1 & 0 & 0 & 0 & 0 &)\cr
(&-1 & 0 & 0 & 0 & 1 & 1 & -1 &)\cr 
(&-1 & 0 & 0 & 1 & 0 &-1 & 1&) \cr  }
\ .\weqn\CVFnullI
$$
leading to the generalized PF system
$$
\eqalign{
\cD_1&=
{{\xt(1)}^2} - \xz(1)\left( 2\xt(1) + \xt(2) + \xt(0) \right) 
   \left( 1 + 2\xt(1) + \xt(2) + \xt(0) \right)
\cr
\cD_2&=
\xt(2)\left( \xt(2) - \xt(0) \right)  - 
  \xz(2)\left( \xt(2) - \xt(0) \right) 
   \left( 2\xt(1) + \xt(2) + \xt(0) \right)
\cr
\cD_0&=
\xt(0)\left( -\xt(2) + \xt(0) \right)  + 
  \xz(0)\left( \xt(2) - \xt(0) \right) 
   \left( 2\xt(1) + \xt(2) + \xt(0) \right)
\ .}\weqn\PFFnullI
$$
The corrections to the mirror map become 
$S_0= -A$, $S_1=-2\, A $, $S_2=-A $, 
where
$$
A=-\sum_{{n_1,n\geq0\atop(n_1,n)\not=(0,0)}}
{(2n_1+2n-1)!\over {n_1!}^2{n!}^2}
 {z_{1}}^{n_1}(z_{0}z_{2})^{n}
\ .\weqn\AFnullI
$$
Moreover the superpotential is given by
$$
W(z_\a,z_0)=\sum_{{n_i\geq0,n_0\not=n_2}}
{\frac{\left(  \xa(0) + 2\xa(1) + \xa(2)-1 \right) !}
   {\xa(0)!{{\xa(1)!}^2}\xa(2)!
     \left( \xa(0) - \xa(2) \right) }}
{z_0}^{n_0} {z_1}^{n_1} {z_2}^{n_2}
\ .\weqn\SPFnullI
$$
The first generating functions for disc instanton numbers
$N_{n_1,n_2}(x)= \sum_{n_0} N_{n_1,n_2,n_0} x^{n_0}$ are
$$
\vbox{
\offinterlineskip\tabskip=0pt\halign{\strut
\vrule~$\ss #$~&\vrule\vrule~$\ss #$~&\vrule~$\ss #$~&\vrule~$\ss #$~\vrule\cr
\noalign{\hrule}
&0&1&2\cr
\noalign{\hrule}\noalign{\hrule}
0
&0 
&-1
&0
\cr
\noalign{\hrule}
1
&{\frc{x}{1 - x}}
&{\frcb{-1 + x + 2\,{x^2}}{1 - x}}
&{\frcb{1 + x - 2\,{x^2} - 3\,{x^3}}{ x-1}}
\cr
\noalign{\hrule}
2
&{\frc{x}{1 - x}}
&{\frcb{1 - 3\,x - 7\,{x^2} + 13\,{x^3} - 6\,{x^4}}
   {{{\left(  x-1 \right) }^3}}}
&{\frcb{2 + 6\,x - 20\,{x^2} - 41\,{x^3} + 46\,{x^4} + 29\,{x^5} - 
     32\,{x^6}}{{{\left(  x-1 \right) }^3}\,\left( 1 + x \right) }}
\cr
\noalign{\hrule}
}}\hfil\weqn\FnullNxI 
$$
The closed formulae for some $N_{n_1,n_2,n_0}$ are:
$$
\eqalign{
N_{0,n_2,n_0}(x)& =0
\ ,n_2>1
\cr
N_{1,n_2,n_0}&=n_2+1
\ ,n_0>n_2+1
\cr
N_{2,0,n_0}& = 
{\Frc{{{\left( 1 + \xa(0) \right) }^2} }{4}}- \coefff14\epsilon (2, \xa(0))
,\forall n_0
\cr
N_{2,1,n_0}& = 
4 + \xa(0) + {{\xa(0)}^2}
\ ,n_0>1
\cr
N_{2,2,n_0}& = 
{\Frcb{45 + 5{{\xa(0)}^2}}{2}} -\coefff12 \epsilon (2, \xa(0))
\ ,n_0>2
\cr
N_{3,0,n_0}& =
{\Frc{{{( 2 + 3\xa(0) +
 {{\xa(0)}^2} ) }^2}}{36}} - 
     \coefff19\epsilon (3, \xa(0))
\ ,\,\forall n_0
\cr
N_{3,1,n_0}& = 
{\Frcb{36 + 14\xa(0) + 17{{\xa(0)}^2} + 4{{\xa(0)}^3} + 
     {{\xa(0)}^4}}{6}}
\ ,n_0>1\ .
\cr
}\weqn\zaspolFnullI
$$
These integers are consistent with results listed in Table 1 of
ref.\ \AVii.

\goodbreak
\subsec{\someline$Y^*=K(F_1)$\someline}
The LSM for D-branes on the canonical bundle of $F_1$ is
defined by the toric polyhedron $\Delop$ with vertices
$$
\eqalign{
\nu_0&=(0,0,0),\ \nu_1=(-1,0,0),\ \nu_2=(1,0,0),\ \nu_3=(1,1,0),\cr
\nu_4&=(0,-1,0),\ \nu_5=(1,1,1),\ \nu_6=(0,0,1).
}
$$
We will first consider a D-brane ending on the outer leg of the toric 
diagram with $|x_1|^2-|x_3|^2=0$. The matrix of charge vectors for this
phase is:
$$
\matrix{\lm 1\cr\lm 2\cr\lm 0}\  = \
\matrix{ 
(&-2&1&1&0&0&0&0&)\cr
(&-1&0&-1&1&1&0&0&)\cr
(&1&0&0&-1&0&1&-1&)}
\ .\weqn\CVFoneIII
$$
The associated Picard-Fuchs operators look:
$$
\eqalign{
\cD_1&=
\xt(1)\left( \xt(1) - \xt(2) \right)  -
  \xz(1)\left( 2\xt(1) + \xt(2) - \xt(0) \right)
   \left( 1 + 2\xt(1) + \xt(2) - \xt(0) \right)
\cr
\cD_2&=\xt(2)\left( \xt(2) - \xt(0) \right)  +
  \xz(2)\left( \xt(1) - \xt(2) \right)
   \left( 2\xt(1) + \xt(2) - \xt(0) \right)
\cr
\cD_0&=
  \left( 2\xt(1) + \xt(2) - \xt(0) \right) \xt(0)-
\xz(0)\left( \xt(2) - \xt(0) \right) \xt(0)
\ .}\weqn\PFFoneIII
$$
The corrections to the mirror map are 
$S_0= A$, $S_1=-2\, A $, $S_2=-A $,
where
$$
A=-\sum_{n_\a\geq0,n_2\leq n_1}
{(-)^{n_2}(2n_1+n_2-1)!\over n_1!(n_2!)^2(n_1-n_2)!}
{z_1}^{n_1} {z_2}^{n_2}
\ .\weqn\AFoneIII
$$
The superpotential \Wouter\ takes the form:
$$
W(z_\a,z_0)=
\sum_{n_\a\geq0,n_0>n_2}
{\frac{{{\left( - \right) }^{\xa(2)}}
     \left(\xa(0) - \xa(2) -1\right) !}{\xa(1)!
     \left( \xa(0) - 2\xa(1) - \xa(2) \right) !
     \left( \xa(1) - \xa(2) \right) !\xa(2)!\xa(0)}}
{z_0}^{n_0} {z_1}^{n_1} {z_2}^{n_2}
\ .\weqn\SPFoneIII
$$
Some generating functions for disc instanton numbers
$N_{n_1,n_2}(x)= \sum_{n_0} N_{n_1,n_2,n_0} x^{n_0}$ are:
$$
\vbox{
\offinterlineskip\tabskip=0pt\halign{\strut
\vrule~$\ss #$~&\vrule\vrule~$\ss #$~&\vrule~$\ss #$~&\vrule~$\ss #$~\vrule\cr
\noalign{\hrule}
&0&1&2\cr
\noalign{\hrule}\noalign{\hrule}
0
&0 
&0
&0 
\cr
\noalign{\hrule}
1
&{\frc{x}{1 - x}}
&{\frc{x\left(  x-2 \right)}{1-x}}
&0
\cr
\noalign{\hrule}
2
&{\frc{{x^3}}{{{\left( 1-x \right) }^3}
\left( 1 + x \right) }}
&{\frc{x\left( -4 + 9x - 7{x^2} + {x^3} \right) }
    {{{\left( 1-x \right) }^3}}}
&{\frc{x\left( -5 + 6x + 5{x^2} - 8{x^3} + {x^5} \right) }
   {{{\left( x-1 \right) }^3}\left( 1 + x \right) }}
\cr
\noalign{\hrule}
}}\weqn\FoneNxIII
$$
With the same conventions as in Table A.1, the
disc instanton numbers of low degree are collected
in the following table.

\def\aa(#1){\hfil{\scriptstyle #1}}
$$
\vbox{
\offinterlineskip\tabskip=0pt\halign{\strut
$#$~\hfil\vrule&
\hfil~$#$~&\hfil~$#$~&\hfil~$#$~&\hfil~$#$~&
\hfil~$#$~&\hfil~$#$~&\hfil~$#$~&\hfil~$#$~&
\hfil~$#$~&\hfil~$#$~
\cr
{\bf 1}&0&1&2&3&4\cr
\noalign{\hrule}
0&          \aa(1)&\aa(  0)&\aa( 0)&\aa( 0)&\aa( 0)\cr
1&          \aa(1)&\aa( -2)&\aa( 0)&\aa( 0)&\aa( 0)\cr
2&          \aa(0)&\aa( -4)&\aa( 5)&\aa( 0)&\aa( 0)\cr
3&          \aa(0)&\aa( -6)&\aa( 35)&\aa( -32)&\aa( 0)\cr
4&          \aa(0)&\aa( -8)&\aa( 135)&\aa( -400)&\aa( 286)\cr
5&          \aa(0)&\aa( -10)&\aa( 385)&\aa( -2592)&\aa( 5187)\cr
6&          \aa(0)&\aa( -12)&\aa( 910)&\aa( -11760)&\aa( 47775)\cr
}}\hskip24pt
\vbox{
\offinterlineskip\tabskip=0pt\halign{\strut
$#$~\hfil\vrule&
\hfil~$#$~&\hfil~$#$~&\hfil~$#$~&\hfil~$#$~&
\hfil~$#$~&\hfil~$#$~&\hfil~$#$~&\hfil~$#$~&
\hfil~$#$~&\hfil~$#$~
\cr
{\bf 2}&0&1&2&3&4\cr
\noalign{\hrule}
0&\aa(0)&\aa( 0)&\aa( 0)&\aa( 0)&\aa( 0)\cr
1&          \aa(1)&\aa( -1)&\aa( 0)&\aa( 0)&\aa( 0)\cr
2&          \aa(0)&\aa( -3)&\aa( 4)&\aa( 0)&\aa( 0)\cr
3&          \aa(0)&\aa( -5)&\aa( 24)&\aa( -21)&\aa( 0)\cr
4&          \aa(0)&\aa( -7)&\aa( 96)&\aa( -261)&\aa( 180)\cr
5&          \aa(0)&\aa( -9)&\aa( 280)&\aa( -1716)&\aa( 3288)\cr
6&          \aa(0)&\aa( -11)&\aa( 684)&\aa( -7956)&\aa( 30604)\cr
}}\hfil
$$
$$
\vbox{
\offinterlineskip\tabskip=0pt\halign{\strut
$#$~\hfil\vrule&
\hfil~$#$~&\hfil~$#$~&\hfil~$#$~&\hfil~$#$~&
\hfil~$#$~&\hfil~$#$~&\hfil~$#$~&\hfil~$#$~&
\hfil~$#$~&\hfil~$#$~
\cr
{\bf 3}&0&1&2&3&4\cr
\noalign{\hrule}
0&\aa(0)&\aa( 0)&\aa( 0)&\aa( 0)&\aa( 0)\cr
1&       \aa(1)&\aa( -1)&\aa( 0)&\aa( 0)&\aa( 0)\cr
2&        \aa(1)&\aa( -4)&\aa( 3)&\aa( 0)&\aa( 0)\cr
3&        \aa(0)&\aa( -8)&\aa( 25)&\aa( -18)&\aa( 0)\cr
4&        \aa(0)&\aa( -12)&\aa( 112)&\aa( -248)&\aa( 153)\cr
5&        \aa(0)&\aa( -16)&\aa( 360)&\aa( -1780)&\aa( 2970)\cr
6&        \aa(0)&\aa( -20)&\aa( 935)&\aa( -8892)&\aa( 29614)\cr
}}\hskip24pt
\vbox{
\offinterlineskip\tabskip=0pt\halign{\strut
$#$~\hfil\vrule&
\hfil~$#$~&\hfil~$#$~&\hfil~$#$~&\hfil~$#$~&
\hfil~$#$~&\hfil~$#$~&\hfil~$#$~&\hfil~$#$~&
\hfil~$#$~&\hfil~$#$~
\cr
{\bf 4}&0&1&2&3&4\cr
\noalign{\hrule}
0& \aa(0)&\aa( 0)&\aa( 0)&\aa( 0)&\aa( 0)\cr
1&        \aa(1)&\aa( -1)&\aa( 0)&\aa( 0)&\aa( 0)\cr
2&        \aa(2)&\aa( -6)&\aa( 4)&\aa( 0)&\aa( 0)\cr
3&        \aa(1)&\aa( -15)&\aa( 34)&\aa( -20)&\aa( 0)\cr
4&        \aa(0)&\aa( -25)&\aa( 168)&\aa( -301)&\aa( 160)\cr
5&        \aa(0)&\aa( -35)&\aa( 580)&\aa( -2349)&\aa( 3385)\cr
6&        \aa(0)&\aa( -45)&\aa( 1600)&\aa( -12584)&\aa( 36288)\cr
}}\hfil
$$
\vskip10pt
\vbox{\leftskip 2pc\rightskip 2pc
\noindent{\ninepoint
{\bf Table A.2:} Disc instanton numbers for the D-brane ending on the outer
edge $|x_1|^2=|x_3|^2$ of $F_1$.}}
\vskip10pt

\ni
Examples for closed formulas for some $N_{n_1,n_2,n_0}$ look:
$$
\eqalign{
N_{n_1,n_2,n_0}& = 0
\ ,n_1<n_2
\cr
N_{1,0,n_0}& = 1
\ ,n_0>1
\cr
N_{1,1,n_0}&= -1
\ ,n_0>2
\cr
N_{2,0,n_0}& = 
{\coefff14{{{\left(  \xa(0)-1 \right) }^2} - \coefff14\epsilon (2, \xa(0))}}
\ , \ \forall n_0
\cr
N_{2,1,n_0}&= 
{\Frcb{-8 + 3\xa(0) - {{\xa(0)}^2}}{2}}
\ ,n_0>1
\cr
N_{2,2,n_0}&=
{\Frcb{15 - 4\xa(0) + {{\xa(0)}^2} }{4} + \coefff14\epsilon (2, \xa(0))}
\ ,n_0>2
\cr
N_{3,0,n_0}&=
\coefff1{36}{{{\left( 2 - 3\xa(0) + {{\xa(0)}^2} \right) }^2}} - 
     \coefff19\epsilon (3, \xa(0))
\ ,\forall n_0
\cr
N_{3,1,n_0}&=
{\Frcb{-108 + 82\xa(0) - 41{{\xa(0)}^2} + 8{{\xa(0)}^3} - 
     {{\xa(0)}^4}}{12}}
\ ,n_0>1
\cr
N_{3,2,n_0}&=
{\Frcb{432 - 194\xa(0) + 71{{\xa(0)}^2} - 10{{\xa(0)}^3} + 
     {{\xa(0)}^4}}{12}}
\ ,n_0>2
\cr
N_{3,3,n_0}&=
{\coefff1{36}({-976\! +\! 348\xa(0)\! -\! 103{{\xa(0)}^2} \!+\!
 12{{\xa(0)}^3}\! - \!
     {{\xa(0)}^4} })\!+\! \coefff19\epsilon (3, \xa(0))}
,n_0>3.
}\weqn\zaspolFoneIII
$$

\goodbreak
\ni
We now consider the flopped Phase with charge vector matrix:
$$
\matrix{\lm 1\cr\lm 2\cr\lm 0}\  = \
\matrix{
(&-2&1&1&0&0&0&0&)\cr
(&0&0&-1&0&1&1&-1&)\cr
(&-1&0&0&1&0&-1&1&)}
\ .\weqn\CVFoneI
$$
The corresponding Picard-Fuchs operators are:
$$
\eqalign{
\cD_1&=
\xt(1)\left( \xt(1) - \xt(2) \right)  -
  \xz(1)\left( 2\xt(1) + \xt(0) \right)
   \left( 1 + 2\xt(1) + \xt(0) \right)
\cr
\cD_2&=
  \xt(2)\left( \xt(2) - \xt(0) \right)+
\xz(2)\left( \xt(1) - \xt(2) \right)
   \left( \xt(2) - \xt(0) \right)
\cr
\cD_0&=
  \xt(0)\left( \xt(0) - \xt(2) \right)+
\xz(0)\left( \xt(2) - \xt(0) \right)
   \left( 2\xt(1) + \xt(0) \right)
\ .}\weqn\PFFoneI
$$
The corrections to the mirror map take the form
$S_0= -A$, $S_1=-2\, A $, $S_2=0$,
with
$$
A=-\sum_{n_\a\geq0,n\leq n_1}
{(-)^{n}(2n_1+n-1)!\over n_1!(n!)^2(n_1-n)!}
{z_1}^{n_1} (z_0z_2)^{n}
\ .\weqn\AFoneI
$$
The superpotential \Winner\ looks:
$$
W(z_\a,z_0)=
\sum_{{n_\a\geq0,n_0>2n_1\atop n_0\not=n_2}}
{\frac{{{\left( - \right) }^{\xa(2)}}
     \left( \xa(0) + 2\xa(1)-1 \right) !}{\xa(0)!\xa(1)!
     \left( \xa(1) - \xa(2) \right) !\xa(2)!
     \left( \xa(0) - \xa(2) \right) }}
{z_0}^{n_0} {z_1}^{n_1} {z_2}^{n_2}
\ .\weqn\SPFoneI
$$
Some generating functions for disc instanton numbers
$N_{n_1,n_2}(x)= \sum_{n_0} N_{n_1,n_2,n_0} x^{n_0}$ are:
$$
\vbox{
\offinterlineskip\tabskip=0pt\halign{\strut
\vrule~$\ss #$~&\vrule\vrule~$\ss #$~&\vrule~$\ss #$~&\vrule~$\ss #$~\vrule\cr
\noalign{\hrule}
&0&1&2\cr
\noalign{\hrule}\noalign{\hrule}
0
&0
&0
&0
\cr
\noalign{\hrule}
1
&{\frc{x}{1 - x}}
&{\frcb{x-1 + {x^2}}{x-1}}
&0
\cr
\noalign{\hrule}
2
&{\frc{x}{{{\left( 1-x \right) }^3}\left( 1 + x \right) }}
&{\frcb{-1 + 3x + 3{x^2} - 8{x^3} + 4{x^4}}
   {{{\left( x-1 \right) }^3}}}
&{\frcb{1 - 4{x^2} - 3{x^3} + 6{x^4} + 3{x^5} - 4{x^6}}
   {{{\left( x-1 \right) }^3}\left( 1 + x \right) }}
\cr
\noalign{\hrule}
3
&{\frc{x\left( 1 + {x^2} \right) }
    {{{\left( 1-x \right) }^5}\left( 1 + x + {x^2} \right) }}
&{\frcb{-1 + 5x + 10{x^2} - 47{x^3} + 65{x^4} - 39{x^5} +
     9{x^6}}{{{\left( x-1 \right) }^5}}}
& *
\cr
\noalign{\hrule}
}}\hfil\weqn\FoneNxIII
$$

Closed formulae for invariants $N_{n_1,n_2,n_0}$ are:
$$
\eqalign{
N_{n_1,n_2,n_0}& = 0
\ ,n_1<n_2
\cr
N_{1,0,n_0}& = 1
\ ,n_0>1
\cr
N_{1,1,n_0}& =-1
\ ,n_0>2
\cr
N_{2,0,n_0}& =
{\coefff14{{\left( 1 + \xa(0) \right) }^2} - \coefff14\epsilon (2, \xa(0))}
\ ,\forall n_0
\cr
N_{2,1,n_0}& =
{-\Frcb{6 + \xa(0) + {{\xa(0)}^2}}{2}}
\ ,n_0>1
\cr
N_{2,2,n_0}& =
{\coefff14({11 + {\xa(0)}^2}) + \coefff14\epsilon (2, \xa(0))}
\ ,n_0>2
\cr
N_{3,0,n_0}& =
{\Frc{{{\left( 2 + 3\,\xa(0) + {{\xa(0)}^2} \right) }^2}}{36}} - 
  {\Frc{\epsilon (3, \xa(0))}{9}}
\ ,\forall n_0
\cr
N_{3,1,n_0}& =
{\Frcb{-60 - 20\,\xa(0) - 23\,{{\xa(0)}^2} - 4\,{{\xa(0)}^3} - 
     {{\xa(0)}^4}}{12}}
\ ,n_0>1
\cr
N_{3,2,n_0}& =
{\Frcb{264 - 2\,\xa(0) + 35\,{{\xa(0)}^2} + 2\,{{\xa(0)}^3} + 
     {{\xa(0)}^4}}{12}}
\ ,n_0>2
\cr
N_{3,3,n_0}& =
{\Frcb{-616 + 54\,\xa(0) - 49\,{{\xa(0)}^2} - {{\xa(0)}^4}}{36}} + 
  {\Frc{\epsilon (3,\xa(0))}{9}}
,n_0>3.
}\weqn\zaspolFoneI
$$
These integers are consistent with the findings of ref.\ \AVii,
listed in their Table 8(I).\foot{We could not reproduce the instanton
numbers given in Table 9 of \AVii\ for the
phase III sketched in their Fig. 19. In fact this table is identical to 
Table 8(II) up to a minus sign and trivial relabelling 
and should simply describe the D-brane on the edge
linearly equivalent to that of phase II.}

\goodbreak
\subsec{\someline$Y^*=K(F_2)$\someline}
The LSM for D-branes on the canonical bundle of $F_2$ is
defined by the toric polyhedron $\Delop$ with vertices
$$
\eqalign{
\nu_0&=(0,0,0),\ \nu_1=(-1,0,0),\ \nu_2=(1,0,0),\ \nu_3=(0,-1,0),\cr
\nu_4&=(2,1,0),\ \nu_5=(0,-1,1),\ \nu_6=(0,0,1).
}
$$
We start with the D-brane ending on the outer leg $|x_1|^2=|x_3|^2$
of the toric diagram, described by the LSM charge vector matrix:
$$
\matrix{\lm 1\cr\lm 2\cr\lm 0}\  = \
\matrix{
(&-2&1&1&0&0&0&0&)\cr
(&0&0&-2&1&1&0&0&)\cr
(&1&-1&0&0&0&1&-1&)}
\ .\weqn\CVFtwoIII
$$
The associated Picard-Fuchs operators look:
$$
\eqalign{
\cD_1&=
\left( \xt(1) - 2\xt(2) \right) \left( \xt(1) - \xt(0) \right)  -
  \xz(1)\left( 2\xt(1) - \xt(0) \right)
   \left( 1 + 2\xt(1) - \xt(0) \right)
\cr
\cD_2&=
-\left( \xz(2)\left( -1 + \xt(1) - 2\xt(2) \right)
     \left( \xt(1) - 2\xt(2) \right)  \right)  + {{\xt(2)}^2}\cr
\cD_0&=
\xz(0)\left( \xt(1) - \xt(0) \right) \xt(0) -
  \left( 2\xt(1) - \xt(0) \right) \xt(0)\ .
\cr
}\weqn\PFFtwoIII
$$
The corrections to the mirror map are $S_0= A$, $S_1=B-2\, A $, $S_2=-2\, B$,
with
$$
\eqalign{
A&=-\sum_{n_1>0,2n_2\leq n_1}{(2
n_1-1)!\over(n_1-2n_2)!{n_2!}^2{n_1}!}
{z_1}^{n_1}{z_2}^{n_2}
\ ,\cr
B&=-\sum_{n_2>0}{(2n_2-1)!\over{n_2!}^2}{z_2}^{n_2}
\ .}\weqn\AFtwoIII
$$
The superpotential \Wouter\ takes the form:
$$
W(z_\a,z_0)=\sum_{{n_\a\geq0,n_0>0\atop n_0>n_1}}
{\frac{{{\left( -1 \right) }^{\xa(1)}}
     \left( \xa(0) - \xa(1) -1\right) !}{\left( \xa(0) -
        2\xa(1) \right) !\left( \xa(1) - 2\xa(2) \right) !
     {{\xa(2)!}^2}\xa(0)}}{z_0}^{n_0} {z_1}^{n_1}
{z_2}^{n_2}
\ .\weqn\SPFtwoIII
$$
\ni The first few generating functions for disc instanton numbers
$N_{n_1,n_2}(x)= \sum_{n_0} N_{n_1,n_2,n_0}
x^{n_0}$ take the form:

$$
\vbox{
\offinterlineskip\tabskip=0pt\halign{\strut
\vrule~$\ss #$~&\vrule\vrule~$\ss #$~&\vrule~$\ss #$~&\vrule~$\ss #$~\vrule\cr
\noalign{\hrule}
&0&1&2\cr
\noalign{\hrule}\noalign{\hrule}
0
&0
&0
&0
\cr
\noalign{\hrule}
1
&x
&x
&0
\cr
\noalign{\hrule}
2
&0
&{\frc{x\left( -3 + x + {x^2} \right) }{ x-1}}
&0
\cr
\noalign{\hrule}
3
&0
&{\frc{x\left( -5 + x + {x^2} + {x^3} \right) }{ x-1}}
&{\frc{x\left( -5 + x + {x^2} + {x^3} \right) }{ x-1}}
\cr
\noalign{\hrule}
}}\hfil\weqn\FtwoNxIII
$$

Some explicit disc instanton numbers $N_{n_1,n_2,n_0}$ are 
(in the same conventions as in Table A.1 ):

\def\aa(#1){\hfil{\scriptstyle #1}}
$$
\vbox{
\offinterlineskip\tabskip=0pt\halign{\strut
$#$~\hfil\vrule&
\hfil~$#$~&\hfil~$#$~&\hfil~$#$~&\hfil~$#$~&
\hfil~$#$~&\hfil~$#$~&\hfil~$#$~&\hfil~$#$~&
\hfil~$#$~&\hfil~$#$~
\cr
{\bf 1}&0&1&2&3&4&5\cr
\noalign{\hrule}
0&\aa(1)&\aa( 0)&\aa( 0)&\aa( 0)&\aa( 0)&\aa( 0)\cr
1&      \aa(1)&\aa( 1)&\aa( 0)&\aa( 0)&\aa( 0)&\aa( 0)\cr
2&      \aa(0)&\aa( 3)&\aa( 0)&\aa( 0)&\aa( 0)&\aa( 0)\cr
3&      \aa(0)&\aa( 5)&\aa( 5)&\aa( 0)&\aa( 0)&\aa( 0)\cr
4&      \aa(0)&\aa( 7)&\aa( 35)&\aa( 7)&\aa( 0)&\aa( 0)\cr
5&      \aa(0)&\aa( 9)&\aa( 135)&\aa( 135)&\aa( 9)&\aa( 0)\cr
6&      \aa(0)&\aa( 11)&\aa( 385)&\aa( 1100)&\aa( 385)&\aa( 11)\cr
}}\hskip25pt
\vbox{
\offinterlineskip\tabskip=0pt\halign{\strut
$#$~\hfil\vrule&
\hfil~$#$~&\hfil~$#$~&\hfil~$#$~&\hfil~$#$~&
\hfil~$#$~&\hfil~$#$~&\hfil~$#$~&\hfil~$#$~&
\hfil~$#$~&\hfil~$#$~
\cr
{\bf 2}&0&1&2&3&4&5\cr
\noalign{\hrule}
0&       \aa(0)&\aa( 0)&\aa( 0)&\aa( 0)&\aa( 0)&\aa( 0)\cr
1&       \aa(0)&\aa( 0)&\aa( 0)&\aa( 0)&\aa( 0)&\aa( 0)\cr
2&       \aa(0)&\aa( 2)&\aa( 0)&\aa( 0)&\aa( 0)&\aa( 0)\cr
3&       \aa(0)&\aa( 4)&\aa( 4)&\aa( 0)&\aa( 0)&\aa( 0)\cr
4&       \aa(0)&\aa( 6)&\aa( 24)&\aa( 6)&\aa( 0)&\aa( 0)\cr
5&      \aa(0)&\aa( 8)&\aa( 96)&\aa( 96)&\aa( 8)&\aa( 0)\cr
6&      \aa(0)&\aa( 10)&\aa( 280)&\aa( 750)&\aa( 280)&\aa( 10)\cr
}}\hfil
$$
$$
\vbox{
\offinterlineskip\tabskip=0pt\halign{\strut
$#$~\hfil\vrule&
\hfil~$#$~&\hfil~$#$~&\hfil~$#$~&\hfil~$#$~&
\hfil~$#$~&\hfil~$#$~&\hfil~$#$~&\hfil~$#$~&
\hfil~$#$~&\hfil~$#$~
\cr
{\bf 3}&0&1&2&3&4&5\cr
\noalign{\hrule}
0& \aa(0)&\aa( 0)&\aa( 0)&\aa( 0)&\aa( 0)&\aa( 0)\cr
1&    \aa(0)&\aa( 0)&\aa( 0)&\aa( 0)&\aa( 0)&\aa( 0)\cr
2&    \aa(0)&\aa( 1)&\aa( 0)&\aa( 0)&\aa( 0)&\aa( 0)\cr
3&    \aa(0)&\aa( 3)&\aa( 3)&\aa( 0)&\aa( 0)&\aa( 0)\cr
4&    \aa(0)&\aa( 5)&\aa( 20)&\aa( 5)&\aa( 0)&\aa( 0)\cr
5&    \aa(0)&\aa( 7)&\aa( 77)&\aa( 77)&\aa( 7)&\aa( 0)\cr
6&    \aa(0)&\aa( 9)&\aa( 225)&\aa( 594)&\aa( 225)&\aa( 9)\cr
}}\hskip25pt
\vbox{
\offinterlineskip\tabskip=0pt\halign{\strut
$#$~\hfil\vrule&
\hfil~$#$~&\hfil~$#$~&\hfil~$#$~&\hfil~$#$~&
\hfil~$#$~&\hfil~$#$~&\hfil~$#$~&\hfil~$#$~&
\hfil~$#$~&\hfil~$#$~
\cr
{\bf 4}&0&1&2&3&4&5\cr
\noalign{\hrule}
0&\aa(0)&\aa( 0)&\aa( 0)&\aa( 0)&\aa( 0)&\aa( 0)\cr
1&     \aa(0)&\aa( 0)&\aa( 0)&\aa( 0)&\aa( 0)&\aa( 0)\cr
2&     \aa(0)&\aa( 1)&\aa( 0)&\aa( 0)&\aa( 0)&\aa( 0)\cr
3&     \aa(0)&\aa( 2)&\aa( 2)&\aa( 0)&\aa( 0)&\aa( 0)\cr
4&     \aa(0)&\aa( 4)&\aa( 16)&\aa( 4)&\aa( 0)&\aa( 0)\cr
5&     \aa(0)&\aa( 6)&\aa( 66)&\aa( 66)&\aa( 6)&\aa( 0)\cr
6&     \aa(0)&\aa( 8)&\aa( 192)&\aa( 512)&\aa( 192)&\aa( 8)\cr
}}\hfil
$$
$$
\vbox{
\offinterlineskip\tabskip=0pt\halign{\strut
$#$~\hfil\vrule&
\hfil~$#$~&\hfil~$#$~&\hfil~$#$~&\hfil~$#$~&
\hfil~$#$~&\hfil~$#$~&\hfil~$#$~&\hfil~$#$~&
\hfil~$#$~&\hfil~$#$~
\cr
{\bf 5}&0&1&2&3&4&5\cr
\noalign{\hrule}
0&     \aa(0)&\aa( 0)&\aa( 0)&\aa( 0)&\aa( 0)&\aa( 0)\cr
1&      \aa(0)&\aa( 0)&\aa( 0)&\aa( 0)&\aa( 0)&\aa( 0)\cr
2&      \aa(0)&\aa( 1)&\aa( 0)&\aa( 0)&\aa( 0)&\aa( 0)\cr
3&      \aa(0)&\aa( 2)&\aa( 2)&\aa( 0)&\aa( 0)&\aa( 0)\cr
4&      \aa(0)&\aa( 3)&\aa( 15)&\aa( 3)&\aa( 0)&\aa( 0)\cr
5&      \aa(0)&\aa( 5)&\aa( 60)&\aa( 60)&\aa( 5)&\aa( 0)\cr
6&      \aa(0)&\aa( 7)&\aa( 175)&\aa( 476)&\aa( 175)&\aa()\cr
}}\hskip25pt
\vbox{
\offinterlineskip\tabskip=0pt\halign{\strut
$#$~\hfil\vrule&
\hfil~$#$~&\hfil~$#$~&\hfil~$#$~&\hfil~$#$~&
\hfil~$#$~&\hfil~$#$~&\hfil~$#$~&\hfil~$#$~&
\hfil~$#$~&\hfil~$#$~
\cr
{\bf 6}&0&1&2&3&4&5\cr
\noalign{\hrule}
0 &    \aa(0)&\aa( 0)&\aa( 0)&\aa( 0)&\aa( 0)&\aa( 0)\cr
1 &     \aa(0)&\aa( 0)&\aa( 0)&\aa( 0)&\aa( 0)&\aa( 0)\cr
2 &     \aa(0)&\aa( 1)&\aa( 0)&\aa( 0)&\aa( 0)&\aa( 0)\cr
3 &     \aa(0)&\aa( 2)&\aa( 2)&\aa( 0)&\aa( 0)&\aa( 0)\cr
4 &     \aa(0)&\aa( 3)&\aa( 16)&\aa( 3)&\aa( 0)&\aa( 0)\cr
5 &     \aa(0)&\aa( 4)&\aa( 60)&\aa( 60)&\aa( 4)&\aa( )\cr
6 &     \aa(0)&\aa( 6)&\aa( 168)&\aa( 477)&\aa( )&\aa( )\cr
}}\hfil
$$
\vskip10pt
\vbox{\leftskip 2pc\rightskip 2pc
\noindent{\ninepoint
{\bf Table A.3:} Disc instanton numbers for the D-brane ending on the outer
edge $|x_1|^2=|x_3|^2=0$ of $F_2$.}}

Closed formulae for some disc instanton numbers $N_{n_1,n_2,n_0}$ are:
$$
\eqalign{
N_{n_1,n_2,n_0}& = 0,
\ n_2\geq n_1,\ (n_1,n_2)\not=(1,1),
\cr
N_{1,0,n_0}& = N_{1,1,n_0}= 0,
\ n_0\not=1
\cr
N_{1,0,1}& = N_{1,1,1}= 1
\cr
N_{n_1,1,n_0}& =
\cases{ n_1-1, & $n_0>n_1$ \cr
        2n_1-n_0, & $1\leq n_0\leq n_1$ \cr
          0, &$n_0=0$ }
\cr
N_{4,2,n_0}& =
{\coefff14(65 - 6n_0 + {{n_0}^2})} - {\coefff14{\epsilon (2,n_0)}}
\ .
}\weqn\zaspolFtwoIII
$$

\goodbreak
\ni
The flop to a phase where the brane has moved to an inner edge
of the toric diagram is described by the charge vectors
$$
\matrix{\lm 1\cr\lm 2\cr\lm 0}\  = \
\matrix{
(&-1&0&1&0&0&1&-1&)\cr
(&0&0&-2&0&1&0&0&)\cr
(&-1&1&0&0&0&-1&1&)}
\ .\weqn\CVFtwoI
$$
The Picard-Fuchs operators look:
$$
\eqalign{
\cD_1&=
\left( \xt(1) - 2\xt(2) \right) \left( \xt(1) - \xt(0) \right)  -
  \xz(1)\left( \xt(1) - \xt(0) \right)
   \left( \xt(1) + \xt(0) \right)
\cr
\cD_2&=
{{\xt(2)}^2}-\xz(2)\left(  \xt(1) - 2\xt(2)-1 \right)
     \left( \xt(1) - 2\xt(2) \right)
\cr
\cD_0&=
\xt(0)\left(\xt(0) -\xt(1)  \right)  +
  \xz(0)\left( \xt(1) - \xt(0) \right)
   \left( \xt(1) + \xt(0) \right)\ .
\cr
}\weqn\PFFtwoI
$$
The corrections to the mirror map become 
$S_0= A$, $S_1=A+B$, $S_2=-2\, B$,
with
$$
\eqalign{
A(z_\a)&=-\sum_{n>0,n_2\geq0}
{(2n-1)!\over(n-2n_2)!{n_2!}^2{n}!}
(z_0z_1)^{n}{z_2}^{n_2}
\ ,\cr
B(z_\a)&=-\sum_{n_2>0}{(2n_2-1)!\over{n_2!}^2}{z_2}^{n_2}
\ .}\weqn\AFtwoIII
$$
The superpotential \Winner\ takes the form:
$$
W(z_\a,z_0)=\sum_{{n_\a\geq0,n_0\not=n_1}}
{\frac{\left(  \xa(0) + \xa(1)-1 \right) !}
   {\xa(0)!\left( \xa(1) - 2\xa(2) \right) !{{\xa(2)!}^2}
     \left( \xa(0) - \xa(1) \right) }}
{z_0}^{n_0} {z_1}^{n_1} {z_2}^{n_2}
\ .\weqn\SPFtwoI
$$

\ni The first generating functions $N_{n_1,n_2}(x)= \sum_{n_0} N_{n_1,n_2,n_0}
x^{n_0}$ \Nx\ for disc instanton numbers are:

$$
\vbox{
\offinterlineskip\tabskip=0pt\halign{\strut
\vrule~$\ss #$~&\vrule\vrule~$\ss #$~&\vrule~$\ss #$~&\vrule~$\ss #$~\vrule\cr
\noalign{\hrule}
&0&1&2\cr
\noalign{\hrule}\noalign{\hrule}
0
&0
&0
&0
\cr
\noalign{\hrule}
1
&-1
&-1
&0
\cr
\noalign{\hrule}
2
&0
&{\frcb{1 + x - 2{x^2} - {x^3}}{x-1}}
&0
\cr
\noalign{\hrule}
3
&0
&{\frcb{1 + x + {x^2} - 3{x^3} - 2{x^4}}{ x-1}}
&{\frcb{1 + x + {x^2} - 3{x^3} - 2{x^4}}{ x-1}}
\cr
\noalign{\hrule}
4
&0
&{\frcb{1 + x + {x^2} + {x^3} - 4{x^4} - 3{x^5}}{x-1}}
&{\frcb{2 \!+ \!2x\! - \!2{x^2} \!+ \!4{x^3}\! - \!30{x^4}
\!-\! 4{x^5} \!+ \!
46{x^6}\! -\! 3{x^7} \!- \!16{x^8}}{{{\left( x -1\right) }^3}
     \left(x -1\right) }}
\cr
\noalign{\hrule}
}}\hfil\weqn\FtwoNxI
$$
\ni Some explicit instanton numbers $N_{n_1,n_2,n_0}$ are:

\def\aa(#1){\hfil{\scriptstyle #1}}
$$
\vbox{
\offinterlineskip\tabskip=0pt\halign{\strut
$#$~\hfil\vrule&
\hfil~$#$~&\hfil~$#$~&\hfil~$#$~&\hfil~$#$~&
\hfil~$#$~&\hfil~$#$~&\hfil~$#$~&\hfil~$#$~&
\hfil~$#$~&\hfil~$#$~
\cr
{\bf 0}&0&1&2&3&4&5\cr
\noalign{\hrule}
0&          \aa(0)&\aa(0)&\aa(0)&\aa(0)&\aa(0)&\aa(0)\cr
1&          \aa(-1)&\aa(- 1)&\aa(0)&\aa(0)&\aa(0)&\aa(0)\cr
2&          \aa(0)&\aa(- 1)&\aa(0)&\aa(0)&\aa(0)&\aa(0)\cr
3&          \aa(0)&\aa(- 1)&\aa(- 1)&\aa(0)&\aa(0)&\aa(0)\cr
4&          \aa(0)&\aa(- 1)&\aa(- 2)&\aa(- 1)&\aa(0)&\aa(0)\cr
5&          \aa(0)&\aa(- 1)&\aa(- 4)&\aa(- 4)&\aa(- 1)&\aa(0)\cr
6&          \aa(0)&\aa(- 1)&\aa(- 6)&\aa(- 11)&\aa(- 6)&\aa(- 1)\cr
}}\hskip15pt
\vbox{
\offinterlineskip\tabskip=0pt\halign{\strut
$#$~\hfil\vrule&
\hfil~$#$~&\hfil~$#$~&\hfil~$#$~&\hfil~$#$~&
\hfil~$#$~&\hfil~$#$~&\hfil~$#$~&\hfil~$#$~&
\hfil~$#$~&\hfil~$#$~
\cr
{\bf 1}&0&1&2&3&4&5\cr
\noalign{\hrule}
0&        \aa(1)&\aa(0)&\aa(0)&\aa(0)&\aa(0)&\aa(0)\cr
1&        \aa(0)&\aa(0)&\aa(0)&\aa(0)&\aa(0)&\aa(0)\cr
2&        \aa(0)&\aa(- 2)&\aa(0)&\aa(0)&\aa(0)&\aa(0)\cr
3&        \aa(0)&\aa(- 2)&\aa(- 2)&\aa(0)&\aa(0)&\aa(0)\cr
4&        \aa(0)&\aa(- 2)&\aa(- 6)&\aa(- 2)&\aa(0)&\aa(0)\cr
5&        \aa(0)&\aa(- 2)&\aa(- 12)&\aa(- 12)&\aa(- 2)&\aa(0)\cr
6&        \aa(0)&\aa(- 2)&\aa(- 20)&\aa(- 40)&\aa(- 20)&\aa(- 2)\cr
}}\hskip15pt
\vbox{
\offinterlineskip\tabskip=0pt\halign{\strut
$#$~\hfil\vrule&
\hfil~$#$~&\hfil~$#$~&\hfil~$#$~&\hfil~$#$~&
\hfil~$#$~&\hfil~$#$~&\hfil~$#$~&\hfil~$#$~&
\hfil~$#$~&\hfil~$#$~
\cr
{\bf 2}&0&1&2&3&4&5\cr
\noalign{\hrule}
0&\aa(0)&\aa(0)&\aa(0)&\aa(0)&\aa(0)&\aa(0)\cr
1&        \aa(0)&\aa(0)&\aa(0)&\aa(0)&\aa(0)&\aa(0)\cr
2&        \aa(0)&\aa(0)&\aa(0)&\aa(0)&\aa(0)&\aa(0)\cr
3&        \aa(0)&\aa(- 3)&\aa(- 3)&\aa(0)&\aa(0)&\aa(0)\cr
4&        \aa(0)&\aa(- 3)&\aa(- 10)&\aa(- 3)&\aa(0)&\aa(0)\cr
5&        \aa(0)&\aa(- 3)&\aa(- 23)&\aa(- 23)&\aa(- 3)&\aa(0)\cr
6&        \aa(0)&\aa(- 3)&\aa(- 40)&\aa(- 88)&\aa(- 40)&\aa(- 3)\cr
}}\hfil
$$
$$
\vbox{
\offinterlineskip\tabskip=0pt\halign{\strut
$#$~\hfil\vrule&
\hfil~$#$~&\hfil~$#$~&\hfil~$#$~&\hfil~$#$~&
\hfil~$#$~&\hfil~$#$~&\hfil~$#$~&\hfil~$#$~&
\hfil~$#$~&\hfil~$#$~
\cr
{\bf 3}&0&1&2&3&4&5\cr
\noalign{\hrule}
0&        \aa(0)&\aa(0)&\aa(0)&\aa(0)&\aa(0)&\aa(0)\cr
1&        \aa(0)&\aa(0)&\aa(0)&\aa(0)&\aa(0)&\aa(0)\cr
2&        \aa(0)&\aa( 1)&\aa(0)&\aa(0)&\aa(0)&\aa(0)\cr
3&        \aa(0)&\aa(0)&\aa(0)&\aa(0)&\aa(0)&\aa(0)\cr
4&        \aa(0)&\aa(- 4)&\aa(- 20)&\aa(- 4)&\aa(0)&\aa(0)\cr
5&        \aa(0)&\aa(- 4)&\aa(- 40)&\aa(- 40)&\aa(- 4)&\aa(0)\cr
6&        \aa(0)&\aa(- 4)&\aa(- 70)&\aa(- 164)&\aa(- 70)&\aa(- 4)\cr
}}\hskip15pt
\vbox{
\offinterlineskip\tabskip=0pt\halign{\strut
$#$~\hfil\vrule&
\hfil~$#$~&\hfil~$#$~&\hfil~$#$~&\hfil~$#$~&
\hfil~$#$~&\hfil~$#$~&\hfil~$#$~&\hfil~$#$~&
\hfil~$#$~&\hfil~$#$~
\cr
{\bf 4}&0&1&2&3&4&5\cr
\noalign{\hrule}
0&      \aa(0)&\aa(0)&\aa(0)&\aa(0)&\aa(0)&\aa(0)\cr
1&      \aa(0)&\aa(0)&\aa(0)&\aa(0)&\aa(0)&\aa(0)\cr
2&      \aa(0)&\aa( 1)&\aa(0)&\aa(0)&\aa(0)&\aa(0)\cr
3&      \aa(0)&\aa( 2)&\aa( 2)&\aa(0)&\aa(0)&\aa(0)\cr
4&      \aa(0)&\aa(0)&\aa(0)&\aa(0)&\aa(0)&\aa(0)\cr
5&      \aa(0)&\aa(- 5)&\aa(- 75)&\aa(- 75)&\aa(- 5)&\aa(0)\cr
6&      \aa(0)&\aa(- 5)&\aa(- 118)&\aa(- 295)&\aa(- 118)&\aa(- 5)\cr
}}\hskip15pt
\vbox{
\offinterlineskip\tabskip=0pt\halign{\strut
$#$~\hfil\vrule&
\hfil~$#$~&\hfil~$#$~&\hfil~$#$~&\hfil~$#$~&
\hfil~$#$~&\hfil~$#$~&\hfil~$#$~&\hfil~$#$~&
\hfil~$#$~&\hfil~$#$~
\cr
{\bf 5}&0&1&2&3&4\cr
\noalign{\hrule}
0&       \aa(0)&\aa(0)&\aa(0)&\aa(0)&\aa(0)&\cr
1&        \aa(0)&\aa(0)&\aa(0)&\aa(0)&\aa(0)&\cr
2&        \aa(0)&\aa( 1)&\aa(0)&\aa(0)&\aa(0)&\cr
3&        \aa(0)&\aa( 2)&\aa( 2)&\aa(0)&\aa(0)&\cr
4&        \aa(0)&\aa( 3)&\aa( 18)&\aa( 3)&\aa(0)&\cr
5&        \aa(0)&\aa(0)&\aa(0)&\aa(0)&\aa(0)&\cr
6&        \aa(0)&\aa(- 6)&\aa(- 210)&\aa(- 600)&\aa(- 210)&\cr
}}\hfil
$$
\vskip10pt
\vbox{\leftskip 2pc\rightskip 2pc
\noindent{\ninepoint
{\bf Table A.4:} Disc instanton numbers for the D-brane ending on the inner
edge $|x_0|^2=|x_3|^2=0$ of $F_2$.}}

Closed formulae for some disc instanton numbers $N_{n_1,n_2,n_0}$ are:
$$
\eqalign{
N_{n_1,n_2,n_0}& = 0,
\ n_2\geq n_1,\ (n_1,n_2)\not=(1,1),
\cr
N_{1,0,n_0}& = N_{1,1,n_0}= 0,
\ n_0>0,
\cr
N_{1,0,0}& = N_{1,1,0}= -1
\cr
N_{n_1,1,n_0}& =
\cases{ n_1-1, & $n_0>n_1$ \cr
           0, &$n_0=n_1$\cr
        -n_0-1, & $0\leq n_0< n_1$ \cr
}
\cr
N_{4,2,n_0}& =
{\coefff14(57 - 2\,\xa(0) + {{\xa(0)}^2})} -
  {\coefff14{\epsilon (2,\xa(0))}}
\ ,n_0>4
\ .
}\weqn\zaspolFtwoI
$$

\listrefs
\end